\documentclass[sigconf, anonymous]{acmart}

\setcopyright{none} % No copyright notice required for submissions
\acmConference[Anonymous Submission to ACM CCS 2017]{ACM Conference on Computer and Communications Security}{Due 19 May 2017}{Dallas, Texas}
\acmYear{2017}

\usepackage{url}
\usepackage{subfigure}
\usepackage{graphicx}
\usepackage{algorithm}
\usepackage{algorithmicx,xspace,algpseudocode}
\usepackage{listings}
\usepackage{pbox}
\usepackage{color}
\usepackage{enumitem}
\makeatletter
\g@addto@macro{\UrlBreaks}{\UrlOrds}
\makeatother
\usepackage{amsmath}
\usepackage{amssymb}

\algrenewcommand{\algorithmiccomment}[1]{$\vartriangleright$ #1}

\makeatletter
\def\old@comma{,}
\catcode`\,=13
\def,{%
  \ifmmode%
    \old@comma\discretionary{}{}{}%
  \else%
    \old@comma%
  \fi%
}
\makeatother

\begin{document}
\title{Exploiting Physical Dynamics to Detect Actuator and Sensor Attacks in Mobile Robots} % TODO: replace with your title

\begin{abstract}
Mobile robots are cyber-physical systems where the cyberspace and the physical world are strongly coupled. Attacks against mobile robots can transcend cyber defenses and escalate into disastrous consequences in the physical world. In this paper, we focus on the detection of active attacks that are capable of directly influencing robot mission operation. Through leveraging physical dynamics of mobile robots, we develop RIDS, a novel robot intrusion detection system that can detect actuator attacks as well as sensor attacks for nonlinear mobile robots subject to stochastic noises. We implement and evaluate a RIDS on Khepera mobile robot against concrete attack scenarios via various attack channels including signal interference, sensor spoofing, logic bomb, and physical damage. Evaluation of 20 experiments shows that the averages of false positive rates and false negative rates are both below 1\%. Average detection delay for each attack remains within 0.40s.
\end{abstract}

% % TODO: replace this section with code generated by the tool at https://dl.acm.org/ccs.cfm
% <concept_id>10002978.10003029.10011703</concept_id>
% <concept_desc>Security and privacy~Usability in security and privacy</concept_desc>

% \ccsdesc{Security and privacy~Use https://dl.acm.org/ccs.cfm to generate actual concepts section for your paper}
% % -- end of section to replace with generated code

% \keywords{template, formatting, pickling} % TODO: replace with your keywords

\maketitle

\section{Introduction}
Recent years have witnessed a rapid growth in the robotics industry. According to International Data Corporation~\cite{idc}, global spending on robotics and related services will reach \$135 billion in 2019. The sheer size of robotics volume is mainly accounted from defense and security, agricultural, medical care, and manufacturing applications~\cite{robotics_stat1}. Recent market predicts a major growth in household and entertainment applications~\cite{akdemir2010emerging}. Mobile robots, as a typical type of robot systems, have capabilities of movement in particular work environments and carry out specific missions. Some representative mobile robots include household cleaning robots such as Roomba, military surveillance drones such as Global Hawk, aerial photography unmanned aerial vehicles (UAV) such as DJI Phantom, Amazon warehouse robots, etc. Major tech companies, e.g., Google, Uber, and Tesla are leading intensive development of autonomous cars to replace human drivers in near future \cite{autotrend}.

The popularity of this emerging technology introduces new security threats to the community. Unlike traditional cyber systems such as computers or mobile phones, mobile robots are characterized by a strong coupling of the cyberspace and the physical world in which they operate. Mobile robots equip sensors, actuators, and electronic control units (ECUs). In a typical control iteration, sensors (e.g., GPS, accelerometer) measure the states (e.g., position, orientation, velocity, etc.) of robots and their surrounding world, and feed the readings to ECUs. ECUs generate control commands based on mission specifications, and actuators (e.g., rotor, wheel) execute them in the physical world. Mobile robots inherit vulnerabilities from their cyber components, and such vulnerabilities can be exploited by adversaries to transcend cyber defenses and further escalate into disastrous consequences in the physical world. Recently, researchers demonstrated several remote hacks into a Jeep Cherokee~\cite{miller2015remote} and Tesla newest models~\cite{teslahack}, and were able to control their actuators such as steering wheels and gas pedals. Moreover, actuators and sensors introduce extra attack surfaces and vulnerabilities into mobile robots. In 2011, an American surveillance drone was brought down by Iranian cyber warfare unit through GPS spoofing attacks~\cite{wiki:iran}. In 2013, a multi-million yacht was demonstrated to be hijacked and controlled using spoofed GPS signals~\cite{zaragoza2013spoofing}. Many missions of robots are safety critical. Hence, it becomes an urgent issue to ensure the security of mobile robots.%Mobile robots are designed for carrying out safety-critical missions, hence, ensuring security becomes an imperative issue.

In this paper, we focus on intrusion detection for mobile robots. We consider attacks that are capable of transcending cyber defenses, actively altering robot behavior and causing damages in the physical world. 
Down to attack consequences, active attacks can be classified into actuator attacks and sensor attacks. \textit{Actuator attacks}, e.g., steering wheel take-over, %are specific to mobile robots, which 
directly alter control commands executed by robot actuators. \textit{Sensor attacks}, e.g., GPS spoofing, alter authentic sensor readings received by controllers. 
%While some studies proposed detection for sensor attacks in linear systems using statistics-based approaches~\cite{sabaliauskaite2016empirical} and behavior-based approaches~\cite{bezzo2014attack, shoukry2015pycra}, sensor attack detection for nonlinear systems is still an open problem.
%Moreover, to the best of the authors' knowledge, {\em{there is no study on the detection of actuator attacks for real world mobile robots.}}
%Given the severe impacts of real-world attacks against mobile robots,
%there is an urgent need to develop a practical IDS which can detect actuator attacks, as well as sensor attacks.

Cyber-layer or cyberspace intrusion detection has been studied extensively in past decades. Traditional host-based~\cite{warrender1999detecting,yeung2003host} and network-based~\cite{roesch1999snort,paxson1999bro} IDSs monitor cyberspace behaviors, e.g., system calls, network events, etc. However, attacks launched through physical channels cannot be detected, since no abnormal cyberspace behavior would be triggered and captured. Wireless sensor network intrusion detection approaches~\cite{krishnamachari2004distributed,li2008pvfs} leverage sensor redundancy to do majority voting on sensor readings and detect inconsistencies between each other. These approaches resort to particular Byzantine thresholds~\cite{castro1999practical} on the number of uncorrupted sensors. When a powerful attack is launched that compromises more sensors than the threshold, the detection fails.

Mobile robots are cyber-physical systems. Beyond the knowledge audited by cyber-layer intrusion detection approaches, they can also access to a {\em second} source of knowledge learned from interacting with the physical world. In particular, the physical dynamics of mobile robots impose constraints on the maneuver of mobile robots. These constraints can be leveraged as a detection vector to provide essential information that reflects ground truth statuses. %These constraints can be leveraged to predict state evolution when attacks are absent. Deviations from such state predictions indicate occurrence of attacks. 
The second source of knowledge is neither obtained nor used in cyber-layer intrusion detection approaches. Noticeably, the information provided by physical dynamics allows for detecting sensor and actuator attacks without resorting to majority voting or Byzantine thresholds.

In this paper, we propose a robot intrusion detection system (RIDS) for nonlinear mobile robots subject to stochastic noises, which leverages the physical dynamics of mobile robots. The proposed RIDS does not assume any sensor or actuator is clean. It is able to detect, pinpoint and quantify sensor and actuator attacks when not all sensors are simultaneously corrupted. The detection capabilities are produced by explicitly leveraging physical dynamics of mobile robots.% To the best of our knowledge, our RIDS is the first one of its kind.

Our main contributions are summarized as follows:
\begin{itemize}[leftmargin=*]
\item To the best of our knowledge, RIDS is the first one that detects sensor and actuator attacks in nonlinear mobile robots subject to stochastic noises.% with unbounded supports.

\item A unique feature provided by RIDS is that it can detect sensor and actuator attacks without resorting to majority voting or Byzantine thresholds.

\item Beyond detection, RIDS is capable of pinpointing attack targets and quantifying attack vectors. The information facilitates intrusion forensics and response.

\item We implement a RIDS for Khepera ground mobile robot and evaluate the RIDS against various concrete attacks. Results indicate that as long as at least 1 out of 3 sensors on Khepera is uncorrupted, the evaluation shows less than 1\% of false positive and false negative detection rate on average. Detection delay remains within an average of 0.40s.
\end{itemize}

\section{Overview}
This section describes the background for general mobile robots and the threat model considered in the paper. For succinctness reason, mobile robots are referred to as robots in the remaining of the paper.%and introduce a running example of a mobile robot mission for experiments and analysis

\begin{figure}[!t]
\centering
\includegraphics[width=\columnwidth]{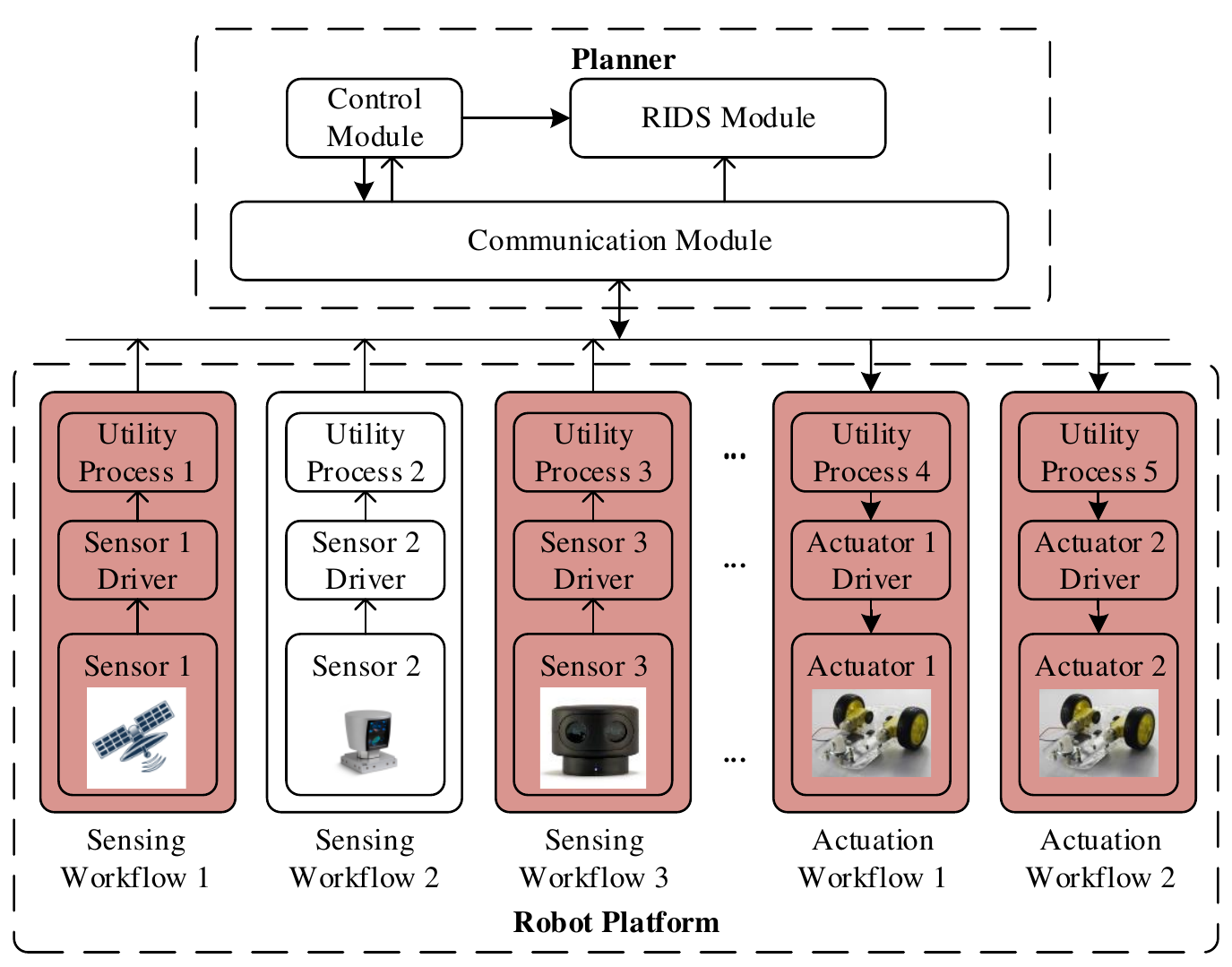}
\caption{Mobile robot system model which consists of a robot platform and a planner. The physical world is perceived by different sensors. Sensor readings go through different processing in utility processes and reach to the planner through a communication module. Control commands are generated from the planner and executed via actuators. (Hollow arrows stand for sensor reading data, and filled arrows stand for control command data.)
}
\label{fig:sys_model}
\end{figure}

\subsection{Sensing and Actuation Workflow}
Figure~\ref{fig:sys_model} shows a general system model for robots. It consists of a robot platform and a planner. The robot interacts with the physical world through sensors and actuators in its physical-layer. Robot cyber-layer runs programs including device drivers, utility processes that perform data processing or translation, etc. We define each sensing procedure from capture of physical signal (e.g., electromagnetic waves, acoustic waves), signal digitization, data processing, to data transmission to the planner as a \emph{sensing workflow}. Analogously, we define the counterpart procedure that receives, translates and executes control commands in an actuator as an \emph{actuation workflow}. Figure~\ref{fig:sys_model} reflects the system model of many real-world robots, such as MIT autonomous car~\cite{leonard2008perception} and Tartan Racing robot~\cite{urmson2007tartan}. %In robots, external sensors provide global information for long-term planning decisions, such as routing. On-board sensors provide local information for short-term safety-related decisions, such collision avoidance and orientation. External sensors and on-board sensors are both essential to robot operation. %A researcher~\cite{maldrone} developed a backdoor program in the control module which allows attackers to remotely control a commodity drone. 
%Son et al.~\cite{son2015rocking} revealed that a resonant frequency of sound can incapacitate a drone through its gyroscope sensor. 

For extensibility and security purposes, recent advances in robot systems adopt modular design principle instead of bulky integration. The development of embedded systems shows a trend of running different tasks of a robot system on separate mission-specific micro-processing chips. For instance, a modern car integrates more than 100 ECUs virtually into every functioning and diagnostics aspect~\cite{koscher2010experimental}. %, and these ECUs are interconnected with control area network(CAN) bus. %The popular Robot Operating System (ROS)~\cite{mcclean2013preliminary} is also built in an isolated design, where different functionality modules run in different processes. 
Microkernels are extensively supported and employed in embedded systems~\cite{de2013microkernel,van2007high} %e.g. ARM Cortex-M cores, 
to keep device drivers and applications isolated by a secure layer. %and constrain abnormally impact within. 
Given these popular design patterns, we model that each sensing workflow or actuation workflow, i.e., device drivers and utility processes, runs in isolation with each other.\label{sec:isolation}

The planner is the control center of a robot. Because of the security and robustness significance of the planner, separation is also enforced between the planner and the robot platform. For instance, the planner could run in a separate chip, or the trust-zone of a processor, or even reside in a physically remote location. It receives sensor readings and sends control commands to the robot platform using certain communication protocols (e.g., CAN)~\cite{di2008understanding}.

% Although at an early stage of robotics development, some works have already revealed security issues and attacks in robots~\cite{checkoway2011comprehensive, denning2009spotlight}. In general, attacks against robots can be categorized into passive attacks and active attacks. Passive attacks eavesdrop and steal sensitive information without intention to sabotage robot missions~\cite{denning2009spotlight}. In contrast, active attacks actively make impacts to robot missions. We only consider active attacks in this work.

In what follows, we depict the attacker and the defender considered in this paper.% Sensor and actuator attacks can be launched through many channels, e.g., sensor spoofing, software logic bomb, message corruption, etc.

\subsection{Threat Model}
% \textbf{Attacker Model}
% The attacker's goal is to cause mission failures or produce damage to the physical world or the robot itself, e.g., collisions with obstacles, taken down, or deviation from original course, etc.

%The attacker will launch active attacks against the robot. Active attacks can be launched through altering sensor readings or control commands. Sensor reading corruptions mislead control algorithms to generate erroneous control commands (sensor attacks). When control commands are corrupted, attacks are able to directly control the robot actuation (actuator attacks).

The attacker considered in this work can observe real-time robot states and has knowledge about robot actuators and sensors. The attacker can launch actuator attacks and/or sensor attacks on one or multiple sensing or actuation workflow(s) through different channels, including malware (e.g., logic bomb), signal interference (e.g., spoofing) or physical damage (e.g., wire cut-off).% Attacks result in corruptions with arbitrary attack vectors on control commands or sensor readings. 

Given the attacker model, our detection system \emph{does not assume any} particular sensing workflow or actuation workflow to be trusted. We assume that an attacker could not compromise all sensing workflows and corrupt all sensor readings simultaneously. Under the design where workflows run with isolation (see Section~\ref{sec:isolation}), the attacker's ability to compromise a workflow does not imply the ability to compromise another. To the best of the authors' knowledge, no reported attack is capable of compromising all sensor workflows and tampering all sensor readings. Admittedly, such attacks could be possible; however, it poses enormous difficulty for attackers. Firstly, for heterogeneous sensors, holding a vulnerability and corresponding exploit targeted on one sensing workflow is costly for an attacker~\cite{2015blackhat,yan2016can}. Hence, it is tough to corrupt all sensors. Secondly, even if an attacker is capable of corrupting all sensors, the attacker needs to launch the attacks simultaneously to avoid detection. It is a great challenge to launch such coordinated attacks on different target sensing workflows.~\cite{petit2015remote}.% Section~\ref{sec:Discussion} discusses more details about the limitation for the detection framework.

The planner contains a defender that aims to detect sensor and actuator attacks targeted on the robot. Because of its security significance, the planner typically maintains minimal code complexity. Its code is extensively tested before deployment, and isolated from the rest of the system. We consider the planner as a trusted computing base (TCB), and the defender keeps no secret information from adversaries.
%With this knowledge, the attacker is capable of sensor reading corruption and/or control command manipulations to achieve his/her attack goal. 
% Actuator attacks and sensor attacks can be launched through different channels. Malware (e.g., Trojan, backdoor, logic bomb, rootkit, exploitable vulnerability, etc.) may hide in any part of the code executing in the robot, e.g., utility libraries, sensor/actuator drivers, etc. For instance, security analysis against ROS shows that robots mounted with ROS are vulnerable to security exploits on multiple software vulnerabilities~\cite{mcclean2013preliminary}. Besides malware, signal interference may intentionally alters the physical, analog signals, such as electromagnetic or acoustic waves received by sensors~\cite{humphreys2008assessing, son2015rocking, shoukry2015pycra, kune2013ghost}. Physical attacks such as intentional sabotage on sensor hardware could also fail the mission. Considering from the perspective of a planner, active attacks can be modeled as data corruptions on sensing or actuation workflow(s), regardless of the attack channel.

\subsection{Our Approach}
In robotics and control theory, \emph{state} refers to the instantaneous description of a dynamical system which changes over time, e.g., the position and orientation of a vehicle, the pitch, yaw, and roll of a drone, etc. %\emph{Input} is the control command generated by ECUs. For instance, for a two-wheel robot, input can be the torque or velocity set for each wheel. 
Control algorithms utilize sensor readings to estimate system states and generate control commands for robot actuators.

\begin{figure}[!t]
\centering
\includegraphics[width=0.55\columnwidth]{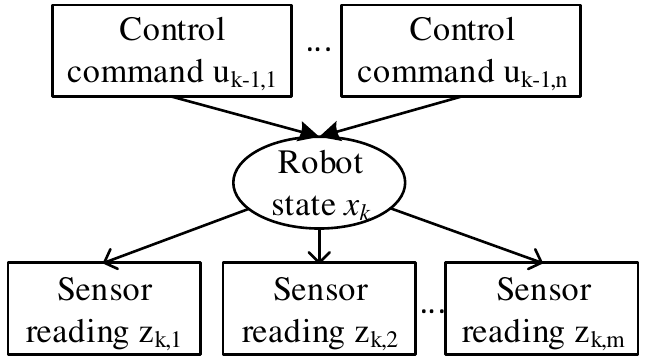}
\caption{Correlation between control commands, robot states, and sensor readings.}\label{fig:determine}
\end{figure}

\textbf{Key insight} In mobile robots, control commands and sensor readings are correlated using robot states as an intermediate (shown in Figure~\ref{fig:determine}). Specifically, executed control commands determine how the robot evolves from an initial state to a new state during a period of time. And the new state is captured by sensor readings. Sensor readings can be utilized to estimate new states. Executed control commands can be estimated through the comparison of the initial and new states. Hence, a discrepancy between planned control commands and executed control commands estimated by sensor readings indicates the existence of actuator attacks. Moreover, multiple sensors in a mobile robot typically have redundancy regarding their measured signals~\cite{cho2004redundancy,chow1984analytical,flynn1985redundant}.
For instance, during a short period, a wheel encoder sensor measures the traveled distance by a wheel, and a LiDAR sensor measures distances between a robot and nearby obstacles. With the knowledge of the robot initial position and heading, both sensors can estimate the current position and heading. Because of sensor redundancy, the states estimated by different sensors could overlap, which can be utilized for detecting sensor attacks by cross-validation. %Specifically, NUISE utilizes sensor readings to estimate robot states as well as executed control commands in each control iteration. 
Therefore, by comparing estimated control commands and planned control commands, we can detect actuator attacks. By comparing estimated states across sensors, we can detect sensor attacks. We develop a RIDS based on this key insight.

\section{Robot Formalization and Problem Statement}\label{sec:sysmodel}
In this section, we formally model the general mobile robot system shown in Figure~\ref{fig:sys_model} and formulate our detection problem. We provide the high level intuition of our approach at the end.

\subsection{Robot Formal Modeling}
A mobile robot can be modeled as a nonlinear discrete time dynamic system. In each control iteration $k\in\{1,2,\cdots\}$, the planner generates planned control commands $\textbf{u}_{k-1}$. After the commands being executed by robot actuators, the robot states evolve from  $\textbf{x}_k-1$ to $\textbf{x}_k$. Under the new states, the planner receives new sensor readings $\textbf{z}_k$. The system model can be formally described by the following equations:
% general discrete time state transition model and sensor model as a
\begin{align}
  \textbf{x}_{k} &= f(\textbf{x}_{k-1},\textbf{u}_{k-1}) + \zeta_{k-1}\nonumber\\
  \textbf{z}_k &= h(\textbf{x}_k) + \xi_k.
  \label{ie011}
\end{align}

The first equation in \eqref{ie011} is referred to as the \textit{kinematic model}, which describes robot state transitions caused by control commands. %Different robots have different configuration of the actuators and different dimension of states. 
The kinematic model specifies the relation between states and control commands based on the actuator properties, e.g., how the actuators function, and where the actuators are located. For instance, a quadcopter's controller adjusts the speeds of the 4 rotors to maneuver itself, while a two-wheel differential drive robot sets different speeds of individual wheels to move along a straight line or take a turn. Function $f(\cdot)$ is referred to as the kinematic function.

The second equation in \eqref{ie011} is the \textit{measurement model}, which describes the relations between sensor readings and robot states. The measurement model is determined by the robot sensor settings, such as sensors types, sensor placement, etc. Function $h(\cdot)$ is referred to as the measurement function. Vectors $\zeta_{k-1}$ represents process noises, which account for external disturbances in the kinematic model. Vectors $\xi_k$ stand for measurement noises, which account for sensing inaccuracy. We assume noise vectors are Gaussian with zero mean and known covariances $Q$ and $R$, respectively. Note that Gaussian noise approximation is common in control system modelings~\cite{kotecha2003gaussian}.%modeling errors and/or all possible random noises such as electrical noise. 

\begin{figure}[!t]
  \centering
  \includegraphics[width =0.85\columnwidth]{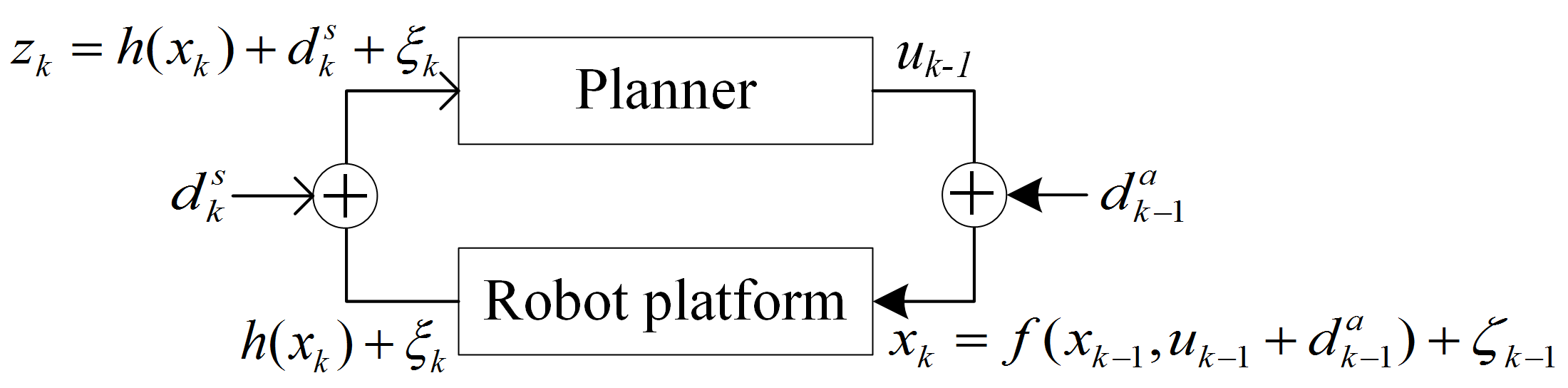}
  \caption{Robot formal modeling considering actuator attacks and sensor attacks.}\label{fig:modeling}
\end{figure}
\begin{figure*}[!ht]
\centering
\includegraphics[width=1.7\columnwidth]{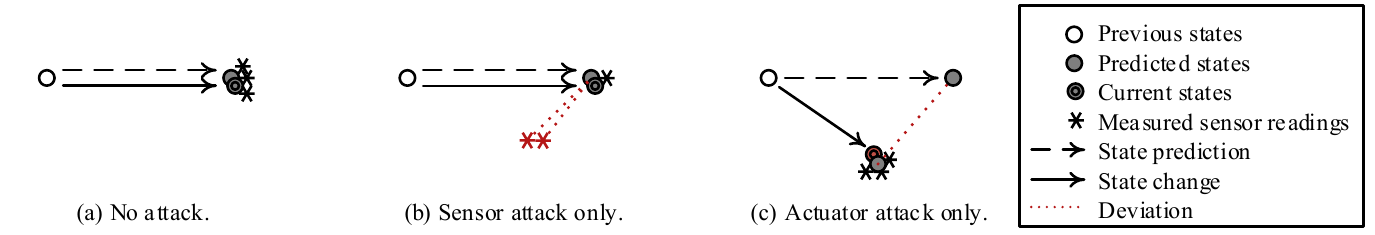}
\caption{High-level intuition for sensor attack and actuator attack detection in robots. The detection is enabled by comparison between measured sensor readings and predicted states. Red circle and red star denotes corrupted states (actuator attack) and corrupted sensor readings (sensor attack), respectively.}
\label{fig:for_the_dummy}
\end{figure*}

System~\eqref{ie011} is general to model all nonlinear robots. Note that the system model is an essential requirement for control purposes during robot design phase. Hence, the modeling described in this section does not introduce extra burden to security administrators.

\textbf{Sensor Attack}\label{sec:se_attack}
 tampers data in a sensor workflow and results in wrong sensor readings received by the planner. When sensor attack is launched, sensor readings $\textbf{z}_k$ received by the planner can be modeled as:
\begin{align}
  \textbf{z}_k &= h(\textbf{x}_k) + \textbf{d}_k^s + \xi_k
  %&\simeq C_k^P\textbf{x}_k + \xi_k^P + a_k^{P}
  \label{eq:un_z_model}
\end{align}
where $\textbf{d}_k^s$ is the attack vector representing corruptions on authentic sensor readings. The robot is free of sensor attack when $\textbf{d}_k^s=0$. Corruptions might exist for multiple sensors. %When attackers have knowledge about the sensor measurement function $h_k(\textbf{x}_k)$, they can carefully craft an attack vector to fulfill their adversarial intent.
%Consider an untrusted positioning system which provides sensor reading as the coordinate of the robot in a 2D sapce: $\textbf{z}_k^{P}=(x_k, y_k)$, and the system uses the position of the robot as the state. We have:
After sensor attacks occur, the control algorithm of the robot might be lured to generate erroneous control commands.

\textbf{Actuator Attack}\label{sec:ac_attack}
 directly alters the control commands executed by the actuators in an actuation workflow. Considering actuator attacks, the kinematic model can be modeled as:
\begin{align}
\textbf{x}_{k} &= f(\textbf{x}_{k-1},\textbf{u}_{k-1}+\textbf{d}_{k-1}^a) + \zeta_{k-1}
% \label{ie019}
\end{align}
%where $G_k$ is an attack input matrix, which specifies the attack dimension (e.g., change left wheel speed only, change angular velocity, etc.); 
where $\textbf{d}_{k-1}^a$ is actuator attack vector. The robot is free of actuator attack when $\textbf{d}_{k-1}^a=0$. %Actuator attacks directly change state transition function of the robot, resulting in undesired behaviors. 
%In some cases, attackers might be capable of completely overriding the control authority of the planner.%To differentiate the two cases, we call them partial control and total control, respectively.

\subsection{Problem Statement}
Consider a robot as modeled in Figure~\ref{fig:modeling} that receives sensor readings $\textbf{z}_k$ from $m$ sensing workflows %$\textbf{z}_k = \{\textbf{z}_{k,1}, \textbf{z}_{k,2}, \cdots, \textbf{z}_{k,m}\}$ 
and sends control commands $\textbf{u}_{k-1}$ to $n$ actuation workflows. %$\textbf{u}_{k-1}=\{\textbf{u}_{k-1,1}, \textbf{u}_{k-1,2}, \cdots, \textbf{u}_{k-1,n}\}$. 
An attacker could launch actuator attack by attack vector $\textbf{d}_{k-1}^a$ and/or launch sensor attack by attack vector $\textbf{d}_k^s$. The robot model with sensor and actuator attacks is:
\begin{align}
  \textbf{x}_{k} &= f(\textbf{x}_{k-1},\textbf{u}_{k-1}+\textbf{d}_{k-1}^a) + \zeta_{k-1}\nonumber\\
  \textbf{z}_k &= h(\textbf{x}_k) + \textbf{d}_k^s + \xi_k
\label{ie019}
\end{align}
In this work, we aim to detect the occurrence of sensor and/or actuator attacks in the robot.. In addition, we intend to identify the specific workflow(s) $i\in\{1,2,\cdots,m\}$ and $j\in\{1,2,\cdots,n\}$ on which attacker targets, and quantify the attack vectors $\textbf{d}_k^s$ and $\textbf{d}_{k-1}^a$ as diagnosis information for future analysis.

\section{High-level Intuition on Why Majority Voting Fails And Our Approach Works}
Under powerful attack scenarios when majority of sensors are corrupted, the IDS's perception of the physical status (e.g., position) of a robot could be greatly distorted. Majority-voting based approaches solely rely on measured sensor readings $\textbf{z}_k$. In our approach, however, the IDS achieves the detection leveraging not only $\textbf{z}_k$ but also physical dynamics. In particular, physical dynamics are leveraged to predict state evolution when attacks are absent. Deviations between such state predictions and measured sensor readings indicate occurrence of attacks. %The fundamental intrusion detection capability is enabled by \emph{the comparison between predicted sensor readings and measured sensor readings}. Unlike traditional voting-based solutions which rely on majority voting among measured values, our approach \emph{does not} require a majority of uncorrupted sensors.

To understand the intuition of our approach, we tentatively consider a robot with 3 sensors at a time instant $k-1$. %Each sensor directly measures the states of the robot, i.e., $\textbf{z}_k = h(\textbf{x}_k)=\textbf{x}_k$. 
For ease of presentation, we tentatively do not consider measurement and process noises. After the execution of control commands, the robot will evolve from the previous states into the current states.

We consider the following possible attack conditions within one control iteration from $k-1$ to $k$.
\begin{itemize}[leftmargin=*]
\item When the robot is free of attack, the 3 measured sensor readings are consistent with each other as shown in Figure~\ref{fig:for_the_dummy} (a). Majority-voting based approaches raise no alarm. In our approach, we firstly predict state evolution. Then we compare measured sensor readings with the predicted states. The consistency indicates that the robot is not under attack. %within ranges determined by sensing and actuation noises.
\item When only sensor attack is launched, and 2 out of 3 sensors are corrupted (Figure~\ref{fig:for_the_dummy} (b)), majority voting-based approaches regard the two sensors that are consistent with each other to be correct and the other one as corrupted. Hence, majority voting makes an obvious mistake here. In our approach, the predicted states serve as the ground truth, and the deviation between predicted states and the 2 measured sensor readings indicates sensor attack. Moreover, our approach can correctly tell which sensors are corrupted and which is not.  
\item When only actuator attack is launched (Figure~\ref{fig:for_the_dummy} (c)), measured sensor readings reflect the actual current states and are consistent with each other, and majority voting-based approaches raises no alarm. On the contrary, in our approach, we notice deviations between the measured sensor readings and the predicted states. The deviations indicate existence of actuator attacks.% After that, the predicted states can be updated using the 3 measured sensor readings, and the deviation between the predicted states and the newly updated predicted states indicates actuator attack vector sizes.% The process of using measured sensor readings to obtain new state estimates is referred to as correction.
% \item When both sensor and actuator attacks are launched (Figure~\ref{fig:for_the_dummy} (d)), majority voting-based approaches make wrong decisions as discussed above. In our approach, we detect actuator attacks first and detect sensor attacks afterwards. However, due to sensor attacks, the measured sensor readings do not agree with each other, and we cannot obtain a consistent predicted states update. To handle this problem, we use each of the 3 sensors to update predicted states, respectively. When a corrupted sensor is used, there will be a larger deviation between the measured sensor readings and the updated predicted states than that using a clean sensor. Therefore, we choose the case that generates smallest deviation, and use the corresponding predicted states to detect sensor attacks for the rest 2 sensors.
\end{itemize}
Based on the intuition, we present the design of proposed robot intrusion detection system in the next section.

\section{Robot Intrusion Detection System Design}\label{sec:rids}
\begin{figure*}[!t]
\centering
\includegraphics[width=2\columnwidth]{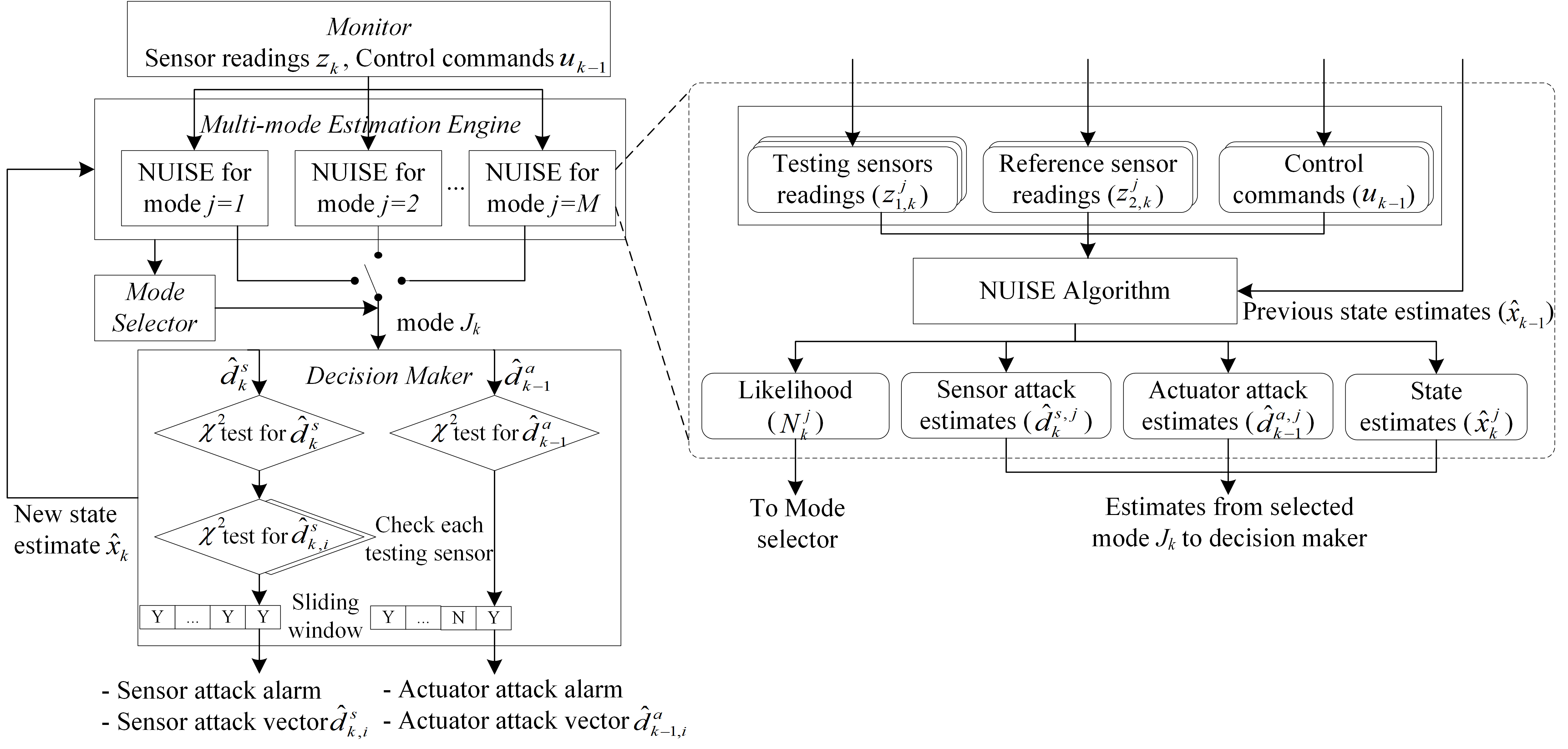}
\caption{Robot intrusion detection system (RIDS) overview.}\label{fig:rids_epl}
\end{figure*}

We develop a robot intrusion detection system (RIDS) framework based on estimation theory for the detection of actuator attacks and sensor attacks. RIDS runs inside the planner (see Figure.~\ref{fig:sys_model}). The defender's goal is as follows: 1) detect actuator and sensor attacks; 2) identify the targets of detected attacks, i.e., which sensing/actuation workflows are attacked; 3) quantify the data corruptions of the detected attacks. %collects sensing data and input in each control iteration, and report attacks to security officer.
% We firstly introduce our RIDS framework, main algorithm, and how it interacts with the rest of the system in Section~\ref{sec:DA}.
% The main algorithm should recursively recall an estimation algorithm to find out the existence of attacks.
% (this paragraph is our solution statement)
% First of all, theoretically, it is important to develop unknown state and input (actuator attack) estimation algorithm in nonlinear systems. This is because the attacks mentioned above cannot be properly detected by the current estimation algorithms. For example, the KF and EKF can only deal with unknown state and the input state estimation algorithm~\cite{SY-MZ-EF:Automatica16} can detect both unknown state and actuator attack only in linear system.
%An application of a estimation theory based attack detection system is practical in robotics because the important requirement for the estimation theory, redundant sensors, are being set up.
%(describe the detection model and interaction with the rest of the system)
%We first introduce an overview of our RIDS framework and how it interacts with the rest of the system. After which, we describe the nonlinear attack detection algorithm.
Figure~\ref{fig:rids_epl} shows the schematic of RIDS. Algorithm~\ref{nalgo1} describes the step-by-step procedure of RIDS. The complete RIDS algorithm is described as Algorithm~\ref{nalgo1_full} in Appendix. Some notations used in the algorithm are explained in Table~\ref{tab:variables}.

RIDS consists of four modules: a monitor, a multi-mode estimation engine, a mode selector, and a decision maker. RIDS runs iteratively from the start until the end of a mission. In each control iteration, the monitor firstly collects data and sends it to the estimation engine. The estimation engine generates a set of estimation results under different hypothesis and their corresponding likelihoods. Then the mode selector accepts the more likely hypothesis. Finally the decision maker leverages the estimation results from the accepted hypothesis to detect attacks. We explain the detail of each module in the sequel.

\begin{table}[!ht]
\centering
\small
\caption{Selected notations in Algorithm~1}
\label{tab:variables}
\begin{tabular}{l|l}
\hline
\textbf{Notation} & \textbf{Explanation}\\
\hline
\hline
$w_s/w_a$ & sliding window size for sensor/actuator attack\\
\hline
$c_s/c_a$ & decision criterion for sensor/actuator attack\\
\hline
$\alpha_s/\alpha_a$ & confidence level for sensor/actuator attack\\
\hline
${\mathcal M}$ & different modes in RIDS\\
\hline
$\textbf{z}_{1,k}^j/\textbf{z}_{2,k}^j$ & testing/reference sensor readings in mode $j$\\
\hline
${\mathcal N}_k^j$ & likelihood of mode $j$\\
\hline
\end{tabular}
\end{table}

\subsection{Monitor}
In each control iteration, control module delivers a copy of control commands $\textbf{u}_{k-1}$ to the monitor of RIDS (Algorithm~\ref{nalgo1} line 2). After control commands execution, the monitor collects sensor readings from all onboard sensors $\textbf{z}_k$ through the communication module (line 3). The monitor sends the received data to the multi-mode estimation engine.
%In Algorithm~\ref{nalgo1}, $\textbf{\hat{x}}_k^P$ is a reduced state vector which estimates 
%Consider $x_k \in {\mathbb R}^{n \times 1 }$ and $d_k \in {\mathbb R}^{m \times 1}$.

\begin{algorithm} \caption{Robot Intrusion Detection System (RIDS)} \label{nalgo1}
\begin{algorithmic}[1]
\Require %Sensor readings $\textbf{z}_k$ from all sensors; control commands $\textbf{u}_{k-1}$ from control module; 
Initial state estimates $\hat{\textbf{x}}_{0|0}$; robot kinematic function $f(\cdot)$; measurement function $h(\cdot)$; parameters $w_s,w_a,c_s,c_a,\alpha_s, \alpha_a$;
\Ensure Detection decision; attack vector estimates
%\State Initialize: $\hat{\textbf{x}}_{0|0}={\mathbb E}[\textbf{x}_0]$; $P_{0}^x=\textsc{P}_0^x$; $P_{-1}^d = \textsc{P}_{-1}^d$
% \State Set 
% \State Initialize;%: $\hat{x}_{0|0}={\mathbb E}[x_0]$; $P_0^x={\mathbb E}[P_0]$; $\mu_k^j=\frac{1}{\mathcal M}$;
% $\tilde{R}_{1,k}=C_{1,0}P_{0}^x C_{1,0}^T + R_{1,0}$;
% $M_{1,0} =(H_{0}^T \tilde{R}_{1,0}^{-1}H_{0})^{-1}H_{0}^T\tilde{R}_{1,0}^{-1}$;
% $P_{0}^{d_1}=(H_{0}^T\tilde{R}_{1,0}^{-1}H_{0})^{-1}$;
% $P_{0}^{xd_1}=(P_{0}^{d_1x})^T=P_{0}^x C_{1,0}^TM_{1,0}^T$; Choose small constant $\epsilon$;
\For{control iteration $k \leftarrow 1$ to $\infty$}
\State Receive control commands $\textbf{u}_{k-1}$;
\State Receive sensor readings $\textbf{z}_k$;
\For{mode $j = 1$ to ${\mathcal M}$}
\State Run NUISE with input ($\textbf{u}_{k-1}$, $\hat{\textbf{x}}_{k-1|k-1}$, $\textbf{z}_{1,k}^j$, $\textbf{z}_{2,k}^j$), 
and generate ($\hat{\textbf{x}}_{k|k}^j$, $\hat{\textbf{d}}_{k}^{s, j}$, $\hat{\textbf{d}}_{k-1}^{a, j}$, ${\mathcal N}_k^j$);
\State $\mu_k^j \leftarrow \max\{{\mathcal N}_k^j \mu_{k-1}^j,\epsilon\}$;
\EndFor
\State Select mode $J_k \leftarrow {\rm argmax}_j$ normalized $\mu_k^j$;
\State Obtain state estimates $\hat{\textbf{x}}_{k|k}^j$, sensor attack estimates $\hat{\textbf{d}}_{k}^{s, j}$, and actuator attack estimates $\hat{\textbf{d}}_{k-1}^{a, j}$ from $J_k$;
\State $b_{k}^s \leftarrow$ Chi-square test on sensor attack;
\State $b_{k}^a \leftarrow$ Chi-square test on actuator attack;
\If{$b_{k}^s =True\; \textbf{and} \;$ sliding window condition met}
\For{each testing sensor $t$ in mode $J_k$}
\State Split sensor attack vector estimates $\hat{\textbf{d}}_{k,t}^s$ from $\hat{\textbf{d}}_{k}^s$;
\If{Chi-square test on $\hat{\textbf{d}}_{k,t}^s=True$}
\State Confirm sensor attack on sensor $t$;
\EndIf
\EndFor
% \State Raise alarm for sensor attack on confirmed sensors;
\EndIf
\If {$b_{k}^{a} =True \; \textbf{and} \;$ sliding window condition met}
     	\State Confirm actuator attack;
        \For{each actuator $t$ in all}
        \State Split actuator attack vector estimates $\hat{\textbf{d}}_{k-1,t}^a$ from $\hat{\textbf{d}}_{k-1}^a$;
        \EndFor
\EndIf\\
%{\footnotemark[\ref{note_algo}]};
%$\|(P_k+P_k^P)^{-\frac{1}{2}}(\textbf{z}_k^{P}-\hat{\textbf{x}}_k)\|_{\infty}>2.58$, then actuator attack exists\footnotemark;
%\State $\|(P_{k-1}^d)^{-\frac{1}{2}}\hat{\textbf{d}}_{k-1}\|_{\infty}>2.58$, then sensor attack exists\footnotemark[\ref{note_algo}];
% \State If $\mathcal{H}_0 : a_k^P = 0$ is rejected{\footnotemark}, then actuator attack exists;
% \State If $\mathcal{H}_0 : d_{k-1} = 0$ is rejected\footnotemark[\ref{note_algo}], then sensor attack exists;
%\hspace*{5.2mm}
\Return Confirmed attack(s); sensor attack estimates $\hat{\textbf{d}}_{k,t}^s$ ($t\in\{\textrm{testing sensors in mode }J_k\}$); actuator attack estimates $\hat{\textbf{d}}_{k-1,t}^a$ ($t\in\{1,\cdots,n\}$);
\EndFor
\end{algorithmic}
\end{algorithm}
% \footnotetext{\label{note_algo}One sample multivariate test with significant level $\alpha$.}
%The framework can be divided into 3 modules: input, NISE, and decision making module. Input module firstly receives trusted sensor readings from the communication module and control inputs from control algorithm module.
%The functionality of the monitor is reflected in Line 4-6 of Algorithm~\ref{nalgo1}.

\subsection{Multi-mode Estimation Engine}\label{sec:NISE}
The goal of the estimation engine is to obtain minimum variance unbiased estimates for actuator attack vectors $\textbf{d}_{k-1}^a$ and sensor attack vectors $\textbf{d}_k^s$ in order to determine attack occurrences. Minimum variance unbiased estimates require that the expected value of estimates should equal to their corresponding target value, and the estimation error variance must be minimized. 
To achieve this goal, we use robot state estimates $\hat{\textbf{x}}_{k}$ as an intermediate, and obtain the attack vector estimates leveraging the correlation between robot states, sensor readings, and control commands as shown in Figure~\ref{fig:determine}. However, the estimation engine faces several challenges.

\textbf{Challenge 1}: Majority of sensors could be potentially corrupted, and we have no knowledge about which sensor(s) is(are) corrupted. Using corrupted sensor readings would result in wrong state and attack vector estimates.

\textbf{Challenge 2}: Existing work does not consider actuator attacks, and directly use planned control commands for state prediction. Under actuator attacks, executed control commands deviate from planned control commands. Since executed control command cannot be directly monitored from the physical world, estimation in the presence of actuator attacks is challenging.

\textbf{Challenge 3}: Real-world robots are nonlinear systems, and they are rooted with inaccuracies in sensing and actuation. It is challenging to build a RIDS which can detect attacks on nonlinear system subject to noises.

To address challenge 1, we propose a multi-mode estimation engine that calculates estimates along with the likelihoods of possible attack conditions. In particular, the multi-mode estimation engine maintains a set of possible sensor attack conditions. Each condition is referred to as a \emph{mode}, which represents a hypothesis that a particular subset of sensors is potentially attacked, and remaining sensors are clean. The potentially attacked sensors are referred to as \emph{testing sensors}, and the clean sensors are referred to as \emph{reference sensors}. Each mode runs a nonlinear unknown input and state estimation (NUISE) algorithm in parallel (line 4-7). Leveraging the reference sensor readings and planned control commands from the last iteration, NUISE estimates new robot states, corruptions on testing sensor readings, corruptions on control commands, and a likelihood for each mode.

\begin{figure*}
\centering
\includegraphics[width=1.7\columnwidth]{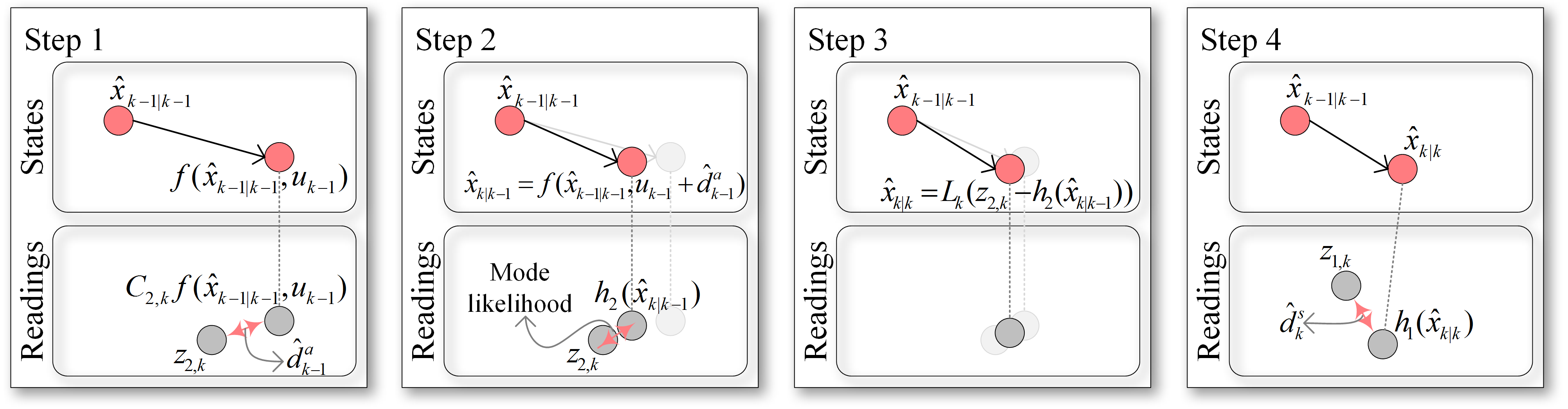}
\caption{Nonlinear unknown input and state estimation algorithm execution. Step 1: actuator attack estimation. Step 2: State prediction with compensation. Step 3: State estimation. Step 4: Testing sensor attack estimation.}
\label{fig:algo_flowchart}
\end{figure*}

\textbf{NUISE algorithm} The NUISE algorithm is described in Figure~\ref{fig:algo_flowchart}. At control iteration $k-1$, the algorithm predicts the states at next iteration using using current state estimates $\hat{\textbf{x}}_{k-1|k-1}$ and planned control commands $\textbf{u}_{k-1}$. The predicted states should reflect a match with the reference sensor readings $\textbf{z}_{2,k}$ in each mode. Whenever a deviation is detected between $\textbf{z}_{2,k}$ and the reflected readings, actuator attacks can be detected (Step 1). With the identified actuator attack estimates from step 1, we conduct a new state prediction with corrected control commands $\textbf{u}_{k-1}+\hat{\textbf{d}}_{k-1}^{a}$ (Step 2). Then the predicted states is corrected by reference sensor readings $\textbf{z}_{2,k}$, and we obtain the state estimates $\hat{\textbf{x}}_{k|k}$ (Step 3). Finally, sensor readings reflected by the state estimates should match all sensor readings, and the deviations between that and testing sensor readings result in the detection of sensor attacks $\hat{\textbf{d}}_{k,t}^{s}$ (Step 4). The full NUISE algorithm is presented as Algorithm~\ref{nalgo2_full} in Appendix.

When a mode is not consistent with the actual attack condition, i.e., corrupted sensors are falsely trusted, reference sensor readings would have a larger discrepancy with state prediction, %Specifically, because of the physical constraint from the kinematic model, the reachable states are limited, and the corrupted sensor readings will be in the unreachable region. 
Subsequently, the state prediction in step 2 cannot be correctly compensated using the actuator attack estimates from step 1. NUISE leverages this discrepancy to generate a likelihood inverse proportional to the discrepancy. 

% Intuitively speaking, when the reference/testing sensor configuration of a mode is incorrect, predicted sensor readings $h_{2}^j(\hat{\textbf{x}}_{k|k-1}^j)$ would result in a relatively larger deviation $\nu_k^j$ with measured reference sensor readings $\textbf{z}_{2, k}^j$, which produces a lower likelihood ${\mathcal N}_k^j$. Hence, ${\mathcal N}_k^j$ reflects the trustworthiness of a mode. Note that control commands $\textbf{u}_{k-1}$ are always regarded as potentially corrupted in each mode, and we test no hypothesis for actuator attacks. Derivation and proof of the NUISE algorithm will be elaborated in Appendix~\ref{sec:AP_n}.
It is a noteworthy point that the proposed detection algorithm does not base on voting mechanism. Even when a majority of sensor readings are corrupted, NUISE generates a higher likelihood for the mode that reflects the ground truth, independent of the number of testing/reference sensors in the mode. The number of modes to be tested grows with the number of sensors in a robot. More information on how to select the mode set is discussed in Section~\ref{sec:Discussion}.

Challenge 2 is also addressed in NUISE algorithm. Using previous state estimates, planned control commands, and reference sensor readings, we calculate the actuator attack vector estimates $\hat{\textbf{d}}_{k-1}^{a,j}$ (Step 1). We compensate the actuator attack vector estimates into the state prediction step (Step 2) to obtain unbiased state prediction.

% As shown in Algorithm~\ref{nalgo2}, leverages robot kinematic model ($\hat{x}_{k|k-1}^j$ in line 7) as a reference during state estimation (line 13). 

% The number of modes to be tested is all possible combinations of testing/reference sensors except the case when all sensors considered as testing sensors. For instance, for a robot with 3 sensors, RIDS tests $2^3-1=7$ hypothesis with different sensor configurations. In the real world, the number of sensors involved in the control loop of a robot is typically small (1-3 sensors). Hence, we believe the number of modes should be within an acceptable amount.

In order to address challenge 3, we model noises with error covariance matrices. The matrices (i.e. noise models) propagation are tracked during each calculation step for all estimation results (see Algorithm~\ref{nalgo2_full} in Appendix~\ref{sec:AP_n}). The matrices serve two purposes: 1) minimizing the variances of the estimates during the estimation process; 2) normalizing attack vector estimates for hypothesis tests. In terms of the nonlinearity of the system, we incorporate nonlinear kinematic and measurement models to minimize estimation error, and use their linearized models to obtain minimum variance estimates. Notice that linearization is performed at the states and controls of each iteration.

\subsection{Mode Selector}
After a normalization, the mode selector compares the likelihood of each mode $\mu_k^j$, and selects the mode $J_k$ with the highest likelihood (line 8). The state and attack vector estimates of the selected mode $J_k$ (line 9) will be leveraged for the decision-making process as follows.

\subsection{Decision Maker}
Using the attack vector estimates $\hat{\textbf{d}}_{k-1}^a$ and $\hat{\textbf{d}}_k^s$, the decision maker conducts Chi-square test to check whether estimated sensor and actuator attack vectors exceed the threshold under a certain level of confidence (line 10-11). In order to reduce the impact of transient fault during the mission, e.g., uneven ground or bump, etc., testing results go through a sliding window and RIDS raises alarm only when a certain number of positives appear in consecutive iterations (line 12 and line 20).

When the number of sensor attack positives exceeds the decision criteria $c_s$, RIDS raises sensor attack alarm. To further confirm that testing sensors are under attack, we separate the sensor attack estimates and conduct Chi-square test separately for an individual testing sensor (line 13-18). %, where $\hat{\textbf{d}}_{k,t}^{s}$ and $P_{k, t}^{s,J_k}$ are the attack vector estimates and the diagonal submatrix of $P_{k}^{s,J_k}$ for testing sensor $t$. 
RIDS reports the confirmed sensors and their corresponding sensor attack vector. Analogously, RIDS raises actuator attack alarm, when the actuator attack positives exceed decision criteria $c_a$ (line 21). RIDS calculates actuator attack vector estimates for each actuator $\hat{\textbf{d}}_{k-1,i}^{a}$. Note that RIDS does not conduct Chi-square test on an individual actuator attack. Instead, it only checks the aggregated test statistics of actuator attack (explained in Appendix~\ref{sec:separation}).

Finally, the decision maker reports confirmed attack(s), sensor attack estimates, and actuator attack estimates to the security administrative as output.

\section{Evaluation}
\begin{figure}[!t]
\centering
\subfigure[Khepera differential drive mobile robot.]{\label{fig:robot}\includegraphics[height=33mm]{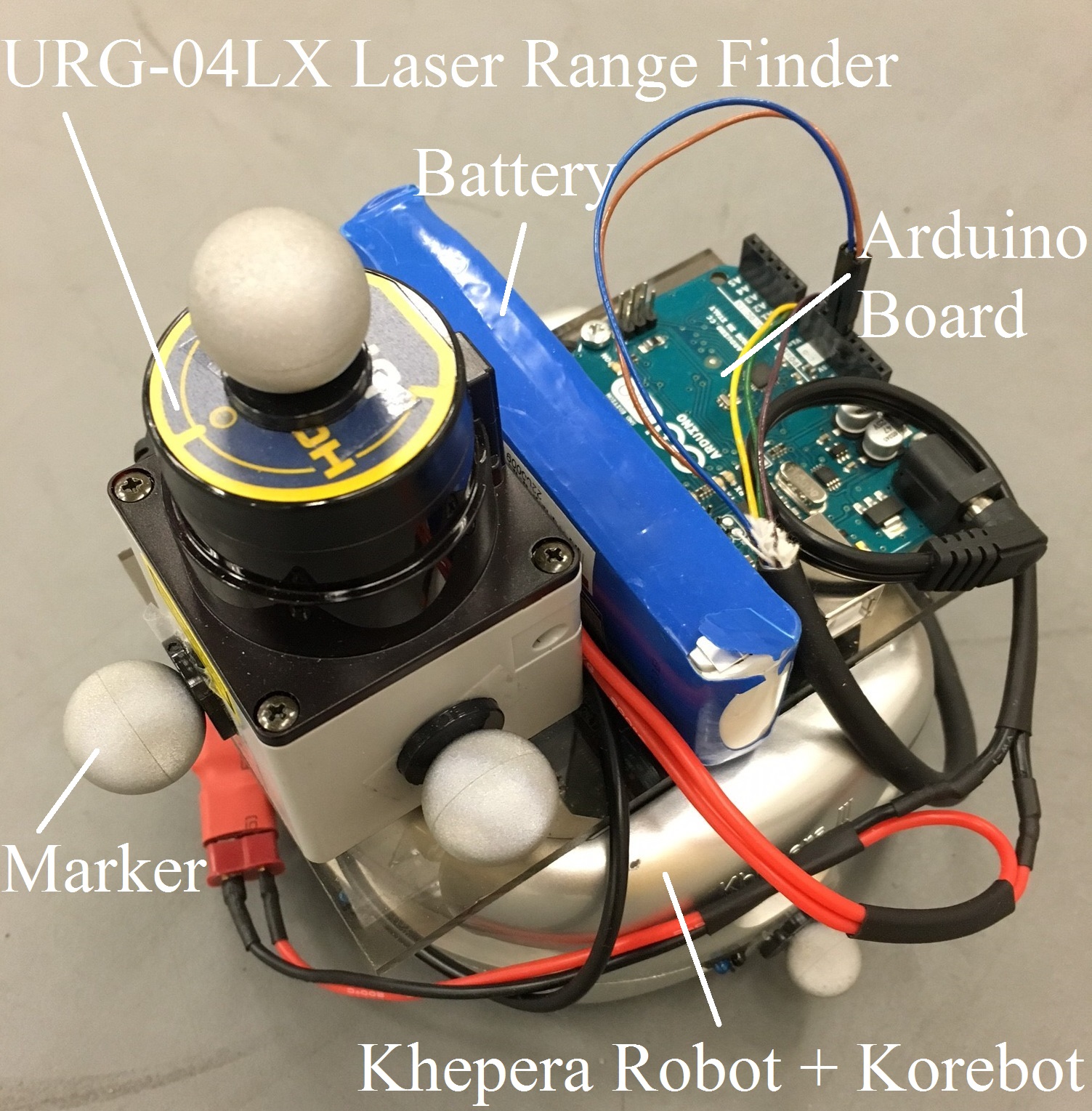}}
% \subfigure[Korebot extension chip mounted Khepera.]{\includegraphics[height=36mm]{khepera_1.JPG}}
\subfigure[Indoor experiment environment with Vicon indoor positioning system.]{\label{fig:env}\includegraphics[height=33mm]{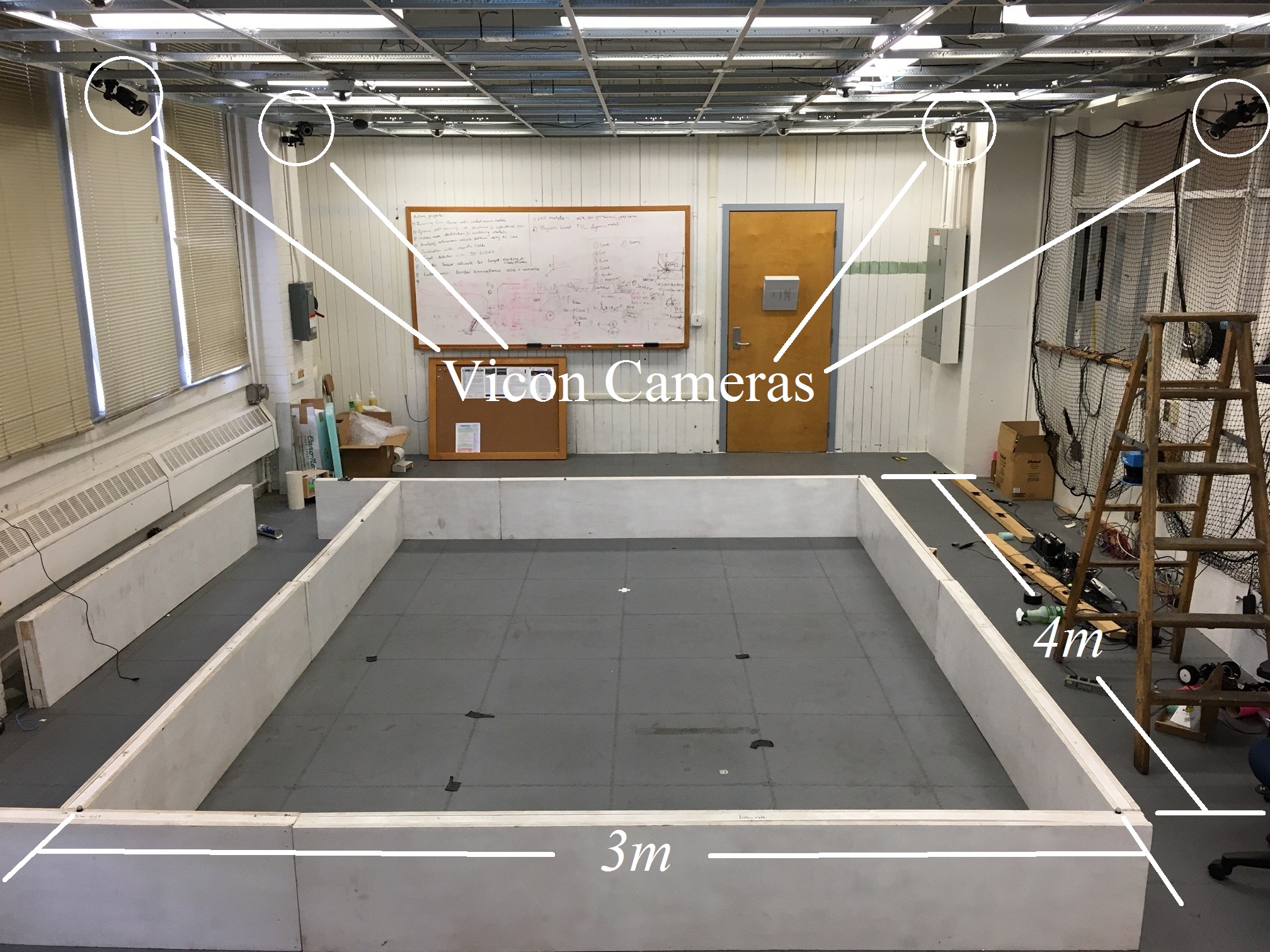}}
% \subfigure[Tracking information in Vicon Tracker v2.2.]{\includegraphics[height=40mm]{vicon2.PNG}}
\caption{Khepera robot testbed and indoor positioning system.}
\label{fig:khepera_korebot}
\end{figure}
To understand the detection effectiveness and efficiency of RIDS for real-world robots, we implement RIDS on a Khepera mobile robot testbed, and conduct experiments under multiple attack scenarios.
%The objective of the evaluation is to answer the following motivating questions: 1) Is RIDS effective to detect actuator and sensor attacks? 2) How efficient does RIDS detect the attacks? 3) Can RIDS quantify the data corruption caused by the attacks? 
In this section, we first introduce the testbed and the mission. Then we describe the experiment setups and attack scenarios launched against Khepera. We analyze the detection results and discuss key parameter selection at last.

\subsection{Robot Platform and Mission}
Figure~\ref{fig:robot} shows an image of the robot system. It consists of Khepera III~\cite{khepera} differential drive robot mounted with KoreBot II~\cite{korebot} extension chip. Khepera is actuated by two wheels on its chassis. KoreBot runs OpenEmbedded Linux, which enables in-robot programming and control. %Another Arduino programming board is placed on top of Khepera and connects with Korebot through USB bus. 
The robot is equipped with three sensors: a wheel encoder, a laser range finder (LiDAR), and an indoor positioning system (IPS). The wheel encoder calculates the traveled distance of each wheel in a short period. Given its previous state, the traveled distance is further processed into its current position and orientation. LiDAR scans laser beams in 240 degrees and receives reflection to obtain distances from surrounding objects. IPS is powered by Vicon motion capturing system (see~\ref{fig:env}), which tracks the position of the robot.
%Multiple cameras on the roof track the positions of the reflective markers on the robot,  
The Kinematic model of Khepera and the measurement models of the three sensors are described in Appendix~\ref{sec:kin_mea_model}.

% In the real world, wheel encoder, ranger finder and positioning system (e.g., GPS) are common sensor settings for many ground vehicles~\cite{hebert2012intelligent, leonard2008perception} in applications such as localization, obstacle avoidance, navigation, etc. Typically. positioning system serves as the primary navigation source, and wheel encoder serves as secondary source when position system is not available (e.g., in tunnels). LiDAR detects nearby obstacles and redirects the robot to avoid them. We believe that our testbed reflects features of real world robots.

%The testbed is consistent with the system model presented in Figure~\ref{fig:sys_model}. The robot consists of three sensing workflows and two actuation workflows. The planner runs in Arduino board and communicates with each sensing and actuation workflow process through TCP socket.

\textbf{Mission} We conduct a motion planning mission where the robot is steered from an initial location to a target location. It avoids collisions with some obstacles on its path. The mission proceeds as follows: 1) Before the mission starts, Khepera receives map information containing the environment setup (obstacles and wall boundaries) and the target location. 2) The planner calculates a collision free path using optimal rapidly-exploring random trees (RRT*) algorithm~\cite{karaman2011sampling}. % The algorithm considers the positions and sizes of the obstacles. 
3) The robot executes PID closed-loop control~\cite{rivera1986internal} to track the planned path using real-time positioning data from IPS.

\subsection{Experiment and Attack Setups}
\begin{table*}[!ht]
\caption{Attack scenarios launched against Khepera mobile robot.}
\label{tab:attack_types}
\centering
\begin{tabular}{l|l}
\hline
\textbf{Attack Scenario} & \textbf{Attack Scenario Description}\\
\hline
\hline
Wheel controller logic bomb & Logic bomb in the actuator utility library that alters the control commands to be executed\\
\hline
Wheel jamming & Physically jamming a particular wheel so that the wheel sticks\\
\hline
IPS logic bomb & Logic bomb in the IPS data processing library that alters authentic positioning data\\
\hline
IPS spoofing & Fake IPS signal that overpowers authentic source and sends out fake positioning data\\
\hline
Wheel encoder logic bomb & Logic bomb in the wheel encoder data processing library that alters readings\\
\hline
LiDAR sensor blocking & Blocking laser ejection and reception in particular angles of the LIDAR\\
\hline
LiDAR DOS & Denial of service by cutting off the LiDAR sensor wire connection\\
\hline
Wheel controller and IPS logic bombs & Altering both wheel control commands and IPS readings through logic bombs\\
\hline
LiDAR DOS and wheel encoder logic bomb & Blocking LiDAR readings and altering wheel encoder readings\\
\hline
IPS spoofing and LiDAR DOS & Altering IPS readings and blocking LiDAR readings\\
\hline
IPS and wheel encoder logic bombs & Altering both IPS and wheel encoder readings through logic bombs\\
\hline
\end{tabular}
\end{table*}

\textbf{Evaluation setup} For comparison purpose, we use an identical path generated from RRT* for all scenarios in the experiments. In each experiment, Khepera travels from a starting point at $(0m, -1.2m)$ to a target point $(0m, 1m)$ inside a $3m\times4m$ confined space shown in Figure~\ref{fig:env}, with constant 7000 speed units\footnote{Speed ratio 144010 units per $m/s$, 7000 units is approximately $0.05m/s$.}. %, and the ground is covered with foam tiles for wheel slippery reduction.
Three $0.8m\times0.2m\times0.2m$ cube-shaped obstacles are on the ground between the starting and the target location. RRT* algorithm generates a path that avoids the obstacles, and Khepera follows the path using PID ($P=0.8, I=0, D=0.001$) control. We identify measurement noise covariance $R$ and the process noise covariance $Q$ by referring to the data sheets of the sensors along with some empirical experiments (refer to~\cite{bavdekar2011identification} for more systematic approaches). RIDS generates detection results under confidence level of $0.05$ for actuator attacks, and $0.005$ for sensor attacks. We choose 2 positives out of 2 windows as the decision criteria for sensor attacks, and choose 3 positives out of 6 windows as the decision criteria for actuator attacks. We will justify how these configurations are chosen in Section~\ref{sec:param_eval} by evaluating RIDS across a range of different parameters.
%To avoid random noise from the environment, such as the uneven ground when moving between foam tiles, we use a sliding window algorithm to process the null-hypothesis testing scores during each iteration. When the number of consecutive scores which are higher than the threshold exceeds the window size, we think an attack occurs. In our case, we use 3 positives out of 3 windows as the criteria for detecting sensor attacks, and use 4 positives out of 5 windows as the criteria for detecting actuator attacks. 

\textbf{Attack setup} We conduct multiple attack scenarios during the mission as described in Table~\ref{tab:attack_types}. We intend to demonstrate that RIDS works well regardless of the attack channels or sensor/actuator targets on the robot. The attack scenarios target on different sensing or actuation workflows of the robot, and launch actuator and sensor attacks through various channels including cyber and physical channels. We inject several logic bombs into the data processing libraries of the IPS and the wheel encoder. The logic bombs can be triggered at a particular time after the mission starts, and continuously alter the authentic sensor readings afterward. For instance, we can trigger the logic bomb to stealthily shift the positioning data received from IPS by a certain distance along the $X$ axis. A logic bomb is also injected into the wheel controller library to alter control commands %or totally override the actuation of the wheels, that is capable of adding extra control commands or totally overriding the actuation of 
for the two wheels. Wheel jamming attack is launched by physically jamming a wheel so that the wheel stops moving. IPS spoofing attack is launched by overriding authentic IPS signals from the Vicon system and sending fake positioning data, analogously to GPS spoofing attacks. For LiDAR, we launch sensor attack by blocking the signal ejection and reception channel in particular directions. Besides, we launch the attack that sabotages the signal transmission by physically cutting off its wire connection. To evaluate the RIDS when multiple sensing workflows or actuation workflows are under attack, we launch several attack scenarios where several of the aforementioned attacks are combined. Table~\ref{tab:attack_setup} shows quantitative information about the details of the attack scenarios. In addition to attack scenarios, we also conduct 9 scenarios under which the mission finishes without attack.

\subsection{Detection Effectiveness}
RIDS aims at detecting, identifying, as well as quantifying attacks in robots. To evaluate the effectiveness of RIDS, we define \emph{true positive} as a time instant that 1) raises alarm if the robot is under attack, and 2) identifies the correct sensor/actuator attack condition, i.e., which sensing or actuation workflow is attacked. Otherwise, positive detection result is considered as \emph{false positive}. \emph{False negative} is defined as a time instant when RIDS does not raise alarm when any workflow is under attack. If all workflows are free of attacks and RIDS does not raise any alarm, the time instant is referred to as \emph{true negative}. The detection result column in Table~\ref{tab:attack_setup} shows identification of attack type and attack condition for different scenarios. From the 11 attack scenarios, we observe that both types of attacks launched from different channels can be successfully detected and identified. Scenario \#1, \#2 and \#8 involves actuator attacks launched from different channels. The robot is under both actuator and sensor attack under scenario \#8. Under scenario \#8, \#9 and \#10, 2 out of 3 sensors on the robot are corrupted and only one sensor remains uncorrupted.
\begin{table*}[!t]
\centering
\small
\caption{Attack scenarios and detection results from RIDS.}
\label{tab:attack_setup}
% \setlength\tabcolsep{1.5pt}
% \def\arraystretch{1.2}
% \hspace*{-0.075\columnwidth}
\begin{tabular}{l|l|l|l|l||l|l|l}
% \begin{tabular}{c|c|c|c|c||c|c|c}
\hline
\textbf{\#} & \textbf{Attack Scenario} & \textbf{\begin{tabular}[l]{@{}l@{}}Launch\\ Time ({\normalfont s})\end{tabular}} & \textbf{\begin{tabular}[l]{@{}l@{}}Attack Type\\(Channel)\end{tabular}} & \textbf{Attack Description} & \textbf{\begin{tabular}[l]{@{}l@{}}Detection\\ Result\end{tabular}} & \textbf{\begin{tabular}[l]{@{}l@{}}Detection\\ Delay ({\normalfont s})\end{tabular}} & \textbf{\begin{tabular}[l]{@{}l@{}}FPR/FNR\protect\footnotemark\end{tabular}}\\ \hline\hline
1 & \begin{tabular}[l]{@{}l@{}}Wheel controller\\logic bomb\end{tabular} & 16.0 & \begin{tabular}[l]{@{}l@{}}Actuator\\(cyber)\end{tabular} & \begin{tabular}[l]{@{}l@{}}-6000 units on $v_L$\\ +6000 units on $v_R$\end{tabular} & A$_{0\rightarrow 1}$\protect\footnotemark & 0.49 & \begin{tabular}[l]{@{}l@{}}A: 0 / 0.83\%\\S: 1\% / -\end{tabular}\\ \hline
%2 & \begin{tabular}[l]{@{}l@{}}Wheel controller\\logic bomb\end{tabular} & 4.0 & \begin{tabular}[l]{@{}l@{}}Actuator\\(cyber)\end{tabular} & (Override\protect\footnotemark) \begin{tabular}[l]{@{}l@{}}7000 units on $v_L$\\ 6500 units on $v_R$\end{tabular} & A & 2.32 & \begin{tabular}[l]{@{}l@{}}A: 0 / 14.2\%\\S: 0 / -\end{tabular}\\ \hline
2 & Wheel jamming & 5.3 & \begin{tabular}[l]{@{}l@{}}Actuator\\(physical)\end{tabular} & \begin{tabular}[l]{@{}l@{}}0 unit on $v_L$\end{tabular} & A$_{0\rightarrow 1}$ & 0.76 & \begin{tabular}[l]{@{}l@{}}A: 0 / 3.1\%\\S: 0 / -\end{tabular}\\ \hline
3 & IPS logic bomb & 19.0 & \begin{tabular}[l]{@{}l@{}}Sensor\\(cyber)\end{tabular} & \begin{tabular}[l]{@{}l@{}}shift $+0.07m$ on X\end{tabular} & \begin{tabular}[l]{@{}l@{}}S$_{0\rightarrow 1}$\end{tabular} & 0.30 & \begin{tabular}[l]{@{}l@{}}A: 0 / -\\S: 1.6\% / 0.24\%\end{tabular}\\ \hline
4 & IPS spoofing & 26.0 & \begin{tabular}[l]{@{}l@{}}Sensor\\(physical)\end{tabular} & \begin{tabular}[l]{@{}l@{}}shift $-0.1m$ on X\end{tabular} & \begin{tabular}[l]{@{}l@{}}S$_{0\rightarrow 1}$\end{tabular} & 0.24 & \begin{tabular}[l]{@{}l@{}}A: 2.24\% / -\\S: 1.55\% / 1.39\%\end{tabular}\\ \hline
5 & \begin{tabular}[l]{@{}l@{}}Wheel encoder\\logic bomb\end{tabular} & 16.0 & \begin{tabular}[l]{@{}l@{}}Sensor\\(cyber)\end{tabular} & \begin{tabular}[l]{@{}l@{}}increment 100 steps on\\ left wheel encoder\end{tabular} &\begin{tabular}[l]{@{}l@{}} S$_{0\rightarrow 2}$\end{tabular} & 0.43 & \begin{tabular}[l]{@{}l@{}}A: 1.4\% / -\\S: 0 / 0.45\%\end{tabular}\\ \hline
6 & LiDAR DOS & 0.0 & \begin{tabular}[l]{@{}l@{}}Sensor\\(physical)\end{tabular} & \begin{tabular}[l]{@{}l@{}}received distance reading is\\ $0m$ in each direction\end{tabular} & \begin{tabular}[l]{@{}l@{}}S$_{3}$\end{tabular} & 0.23 & \begin{tabular}[l]{@{}l@{}}A: 0 / -\\S: 0 / 0\end{tabular}\\ \hline
7 & \begin{tabular}[l]{@{}l@{}}LiDAR sensor\\blocking\end{tabular} & 7.0 & \begin{tabular}[l]{@{}l@{}}Sensor\\(physical)\end{tabular} & \begin{tabular}[l]{@{}l@{}}received distance reading to\\ the left wall is incorrect\end{tabular} & \begin{tabular}[l]{@{}l@{}}S$_{0\rightarrow 3}$\end{tabular} & 0.55 & \begin{tabular}[l]{@{}l@{}}A: 0.22\% / -\\S: 0 / 0.80\%\end{tabular}\\ \hline
8 & \begin{tabular}[l]{@{}l@{}}Wheel controller \&\\IPS logic bomb\end{tabular} & \begin{tabular}[l]{@{}l@{}}W: 10.0\\I: 3.8\end{tabular} & \begin{tabular}[l]{@{}l@{}}Sensor\&Actuator\\(cyber)\end{tabular} & \begin{tabular}[l]{@{}l@{}}$\mp$6000 units on $v_L$, $v_R$\\shift $+0.07m$ on X\end{tabular} & \begin{tabular}[l]{@{}l@{}}A$_{0\rightarrow 1}$\\S$_{0\rightarrow 1}$\end{tabular} & \begin{tabular}[l]{@{}l@{}}W: 0.59\\I: 0.50\end{tabular} & \begin{tabular}[l]{@{}l@{}}A: 0 / 1.8\%\\S: 0 / 0.24\%\end{tabular}\\
\hline
9 & \begin{tabular}[l]{@{}l@{}}LiDAR DOS \&\\wheel encoder\\logic bomb\end{tabular} & \begin{tabular}[l]{@{}l@{}}W: 16.0\\L: 25.0\end{tabular} & \begin{tabular}[l]{@{}l@{}}Sensor\\(cyber\&physical)\end{tabular} & \begin{tabular}[l]{@{}l@{}}increment 100 steps on left wheel\\$0m$ in each direction from LiDAR\end{tabular} & \begin{tabular}[l]{@{}l@{}}S$_{0\rightarrow2\rightarrow4}$\end{tabular} & \begin{tabular}[l]{@{}l@{}}W: 0.43\\L: 0.29\end{tabular} & \begin{tabular}[l]{@{}l@{}}A: 0 / -\\S: 0.48\% / 0.72\%\end{tabular}\\
\hline
10 & \begin{tabular}[l]{@{}l@{}}IPS spoofing \&\\LiDAR DOS\end{tabular} & \begin{tabular}[l]{@{}l@{}}L: 10.0\\I: 17.0\\L: 25.0\end{tabular} & \begin{tabular}[l]{@{}l@{}}Sensor\\(physical)\end{tabular} & \begin{tabular}[l]{@{}l@{}}$0m$ in each direction from LiDAR\\shift $+0.07m$ on X\\LiDAR readings are restored to normal\end{tabular} & \begin{tabular}[l]{@{}l@{}}S$_{ 0\rightarrow3\rightarrow 5\rightarrow1}$\end{tabular} & \begin{tabular}[l]{@{}l@{}}L: 0.36\\I: 0.29\\L: 0.30\end{tabular} & \begin{tabular}[l]{@{}l@{}}A: 0.25\% / -\\S: 0.25\% / 0.58\%\end{tabular}\\
\hline
11 & \begin{tabular}[l]{@{}l@{}}IPS \&\\wheel encoder\\logic bomb\end{tabular} & \begin{tabular}[l]{@{}l@{}}W: 10.0\\I: 28.0\end{tabular} & \begin{tabular}[l]{@{}l@{}}Sensor\\(cyber)\end{tabular} & \begin{tabular}[l]{@{}l@{}}increment 100 steps on left wheel\\shift $+0.1m$ on X\end{tabular} & \begin{tabular}[l]{@{}l@{}}S$_{0\rightarrow2\rightarrow6}$\end{tabular} & \begin{tabular}[l]{@{}l@{}}W: 0.33\\I: 0.31\end{tabular} & \begin{tabular}[l]{@{}l@{}}A: 0 / -\\S: 0.25\% / 0.33\%\end{tabular}\\
\hline
\end{tabular}
\end{table*}
\addtocounter{footnote}{-1}
\footnotetext{False positive rate and false negative rate.}
\addtocounter{footnote}{1}
\footnotetext{Subscript ${i\rightarrow j}$ stands for transition from sensor/actuator mode $i$ to mode $j$. W, I, and L stands for wheel encoder, IPS, and LiDAR, respectively.}
% \addtocounter{footnote}{1}
% \footnotetext{Override the control commands and execute arbitrary control commands.}

For the ease of presenting classification results, Table~\ref{tab:modes} defines the possible attack conditions for actuator and sensor attacks. We refer to these attack conditions as sensor modes and actuator modes. % Note that sensor attack condition S$_4$, S$_5$ and S$_6$ describes when multiple sensor readings are corrupted and only one sensor returns uncorrupted value. In each control iteration, we also calculate the deviation between reference sensor readings and estimated sensor readings using state estimates so that we see the estimation for all sensors.
\begin{table}[!ht]
\centering
% \footnotesize
\caption{Sensor and actuator mode definition.}
\label{tab:modes}
\begin{tabular}{l|l}
\hline
\textbf{\begin{tabular}[l]{@{}l@{}}Sensor\\Mode \#\end{tabular}} & \textbf{Robot Attack Condition}\\
\hline
S$_0$ & under no sensor attack \\
\hline
S$_1$ & under IPS sensor attack \\
\hline
S$_2$ & under wheel encoder sensor attack \\
\hline
S$_3$ & under LiDAR sensor attack \\
\hline
S$_4$ & under wheel encoder and LiDAR sensor attack \\
\hline
S$_5$ & under IPS and LiDAR sensor attack \\
\hline
S$_6$ & under IPS and wheel encoder sensor attack \\
\hline
\hline
\textbf{\begin{tabular}[l]{@{}l@{}}Actuator\\Mode \#\end{tabular}} & \textbf{Robot Attack Condition}\\
\hline
A$_0$ & under no actuator attack \\
\hline
A$_1$ & under actuator attack \\
\hline
\end{tabular}
\end{table}
Figure~\ref{fig:rids_est_prob} presents graphical details of the detection results for several attack scenarios. Each figure includes eight plots that depicts the outputs from RIDS: 1) IPS sensor attack vector estimates ($\textbf{d}_{k,I}^s$); 2) wheel encoder sensor attack vector estimates ($\textbf{d}_{k,W}^s$); 3) LiDAR sensor attack vector estimates ($\textbf{d}_{k,L}^s$); 4) actuator attack vector estimates for the wheels ($\textbf{d}_{k}^a$); 5) sensor attack Chi-square hypothesis test statistic and threshold under confidence level $\alpha=0.005$; 6) sensor mode selection; 7) actuator attack Chi-square hypothesis test statistic and threshold under confidence level $\alpha=0.05$; 8) actuator mode selection. Figure~\ref{fig:est_prob_a18} shows a scenario when wheel controller control commands and IPS sensor readings are tampered by logic bombs at different time instants. Around $4s$, IPS sensor attack vector estimates on the X axis surge (plot 1). Accordingly, sensor attack test statistic surges above the threshold (plot 5), and sensor mode selection (plot 6) indicates that the robot is under IPS sensor attack. Around $10s$, actuator attack vector estimates on the left and right wheel significantly deviate from 0. Accordingly, we notice an oscillating surge over the threshold for actuator attack (plot 7), and actuator mode selection (plot 8) indicates that the robot is under actuator attack. Throughout the experiment, both sensor attack estimates for wheel encoder and LiDAR remain silent. Figure~\ref{fig:est_prob_a5} shows a scenario where attacks against multiple sensors are launched/revoked at four different time instants. We observe that the detection results are highly consistent with the attack scenario. Detection results for some other scenarios can be found in Appendix~\ref{sec:more_figures}.

We examine the false positive and false negative time instants occurred in the experiments. Majority of false classifications are introduced by the sliding window for the purpose of transient fault tolerance. False positives and false negatives are inevitable at the edge when an attack becomes active or revoked, and the choice of window size and decision criteria determines the number of false classifications. For sensor attack false positives, we observe only a small portion is caused by sensor or actuator mode selection errors, while the majority is caused by bogus test statistics increases. The average false positive rate and false negative rates are 0.86\% and 0.97\%, respectively. Therefore, we believe the RIDS can be considered as effective in detecting and identifying both actuator attacks and sensor attacks targeted on our testbed.% We notice that majority of false positives and false negatives are caused by noises during sudden direction tuning or running across uneven ground tiles.

\subsection{Detection Delay}\label{sec:delay}
Detection delay indicates the time between when an attack is launched/revoked, and when RIDS captures the change. Theoretically, in each control iteration, attack vectors can be revealed in the very next iteration after launch. However, we add a sliding window in the decision maker to eliminate transient fault impact. Hence, detection delays will depend on the parameter choice. In our experiment, we choose 2/2 and 3/6 as the decision criteria and sliding window size. The detection delay for each attack scenario is shown in Table~\ref{tab:attack_setup}. We observe that the detection delays are quite small. Specifically, average detection delay for sensor attacks is $0.35s$, and the counterpart for actuator attacks is $0.61s$. The average delays are consistent with our parameter selection for actuator and sensor attacks. Through our analysis of the detection statistics, we notice that the test statistics raises above the threshold mostly in the next iteration after an attack occurs. Most delays are incurred by the sliding window decision making.

Once the magnitude of an attack exceed predetermined threshold, the maximal detection delay is a constant multiple of control iterations. The frequency of the control iteration is determined by hardware configurations (e.g., CPU frequency) and control algorithm design, which is chosen to meet the specifications of robots and mission requirement. Fast moving robots have higher frequency of control cycles, hence the detection delay would be small. %For instance, to operate in a harsh field environment under a relatively high speed, Boston Dynamics BigDog~\cite{bigdog} is designed with a frequency of 200HZ to facilitate robot balancing and steering. In our testbed, the frequency of the control iteration is 100HZ. We observe that the detection is much earlier than the collision. Security administrative has plenty of time for attack response. Moreover, since certain safety policies are usually enforced into a mission (e.g., safety minimum distance from obstacles for ground vehicles, safety minimum altitude for UAVs), we believe that our RIDS can quickly detect attacks before they cause significant damage to the robot or the environment.

\subsection{Attack Vector Quantification}
Actuator attack and sensor attack vector estimates provide quantitative information about the attacks, which can assist security administrative for further diagnosis and attack response. For instance, after sensor attack detection in scenario \#9, IPS sensor attack vector estimates on X axis $d_{k, I}^{s, x}$ is $+0.069m$ with a standard deviation of $\pm0.002m$. Average error between estimated vector and the ground truth ($+0.07m$) is $1.91\%$. After actuator attack detection, average actuator attack vector estimates on the left wheel and right wheel are $d_k^{a, L}=-5975.4\pm1188$ units and $d_k^{a, R}=+5892.4\pm1091$ units, respectively. Average error between the estimated vector and the ground truth ($\mp6000$ units) are $0.41\%$ and $1.79\%$, respectively. We observe that the estimation results are fairly accurate for both actuator and sensor attack vector estimates.% We will analyze how quantification accuracy can be improved by adding more sensors in Section~\ref{sec:multi_sensor}.%The reason why sensor attack vector estimates are more accurate is that sensor readings follow physical rules and are governed by certain stochastic restrictions. However, actuator attack vectors have no such restrictions.

\subsection{Parameter selection}\label{sec:param_eval}
We evaluate the detection effectiveness of RIDS across different choices of detection window sizes ($w$), detection criteria ($c$), and detection confidence level ($\alpha$) in detection of actuator and sensor attacks. The analysis is conducted over the 20 experiments including 11 different attack scenarios and 9 no-attack scenarios. Figure~\ref{fig:a_roc_alpha} depicts the ROC curve for actuator attack detection under different confidence levels range from $\alpha=0.0005\sim0.995$. From the figure, we notice that the detection achieves an acceptable performance when $\alpha=0.05$ under different $w$ and $c$ settings. The selection of $w$ and $c$ eliminates the impact of faults during the mission and determines whether a positive time instant should be regarded as an attack. with a chosen $\alpha$, Figure~\ref{fig:a_roc_wc} depicts the detection performance under different $w$ and $c$. The results indicate that under certain window size, detection performance increases first and reduces afterward. We select $c/w=3/6$ as the configuration, which yields the best performance. Analogously, we select $\alpha=0.005$ as the optimal confidence level, and $c/w=2/2$ as the optimal decision criteria/window size configuration for sensor attack detection.

\subsection{Evasive attacks} 
An attacker's ideal goal is to bypass the detection of RIDS, yet be capable of causing significant impact to the robot or the environment it operates in. We can think of two possible ways of crafting such evasive attacks: 1) reduce attack vectors so that the test statistics in RIDS do not raise alarms; 2) frequently switch attack targets so that sliding window will treat the attack vectors as faults. Under the current RIDS configuration ($\alpha$, $w$, $c$ and sensor accuracy) in our experiments, the attack vector needs to be extremely small to remain alarm silence. For instance, we find that the distance shift for IPS sensor attack needs to remain under $0.02m$ to avoid detection. The speed alteration needs to remain under $900$ units ($0.006m/s$) to avoid detection. Moreover, the control algorithm ensures that attack impact does not accumulate as time goes. Hence, we believe that an attacker cannot make a significant impact with the first approach. Since we demonstrate that the detection delay is small in Section~\ref{sec:delay}, the impact of the second attack cannot succeed either.

\section{Discussion}\label{sec:Discussion}
In this section, we discuss some issues related to applying the RIDS to real-world robots.%, and limitation in term of detectability.

\textbf{General applicability}\label{sec:uav} RIDS is applicable to nonlinear systems, which covers majority of real-world complicated robot systems, such as UAV~\cite{nemra2009robust} and quadrotor~\cite{kumar2012opportunities}.
The design and implementation of RIDS only require the kinematic model and the measurement model. In fact, both models are the essential requirements for any robot mission. Therefore, the RIDS incurs little extra mathematical modeling burden for security administrative. We present how RIDS can be applied to UAV as another type of nonlinear system in Appendix~\ref{sec:app_uav}.
%In this paper, we use Khepera, a ground mobile robot, as a running example to demonstrate how to apply the NUISE algorithm and its effectiveness in terms of sensor and actuator attack detection. 

% For actuator attack against the rotors in UAV, attack vector in~\eqref{ie019} can be represented as $\textbf{d}_k^a \triangleq [d^x,d^y,d^z,d^{\phi},d^{\theta},d^{\psi}]^T$. % with matrix $G_k$ such that $G_k \textbf{d}_k = [0,0,0,d^x,d^y,d^z,d^{\phi},d^{\theta},d^{\psi}]^T$.
% For sensor attacks that modify GPS signals of $[x_k,y_k,z_k]^T$, the GPS measurement model is in~\eqref{eq:un_z_model} can be represented as 
% \begin{align*}
% h_k^u(\textbf{x}_k) =
% \left[
% \begin{array}{ccccccccc}
% 1&0&0&0&0&0&0&0&0\\
% 0&1&0&0&0&0&0&0&0\\
% 0&0&1&0&0&0&0&0&0\\
% \end{array}
% \right]
% \end{align*}
% and sensor attack vector $a_k^u = [a_k^{x},a_k^{y},a_k^{z}]^T$.

\textbf{Mimicry attacks}
Admittedly, RIDS cannot handle all active attacks targeted on robots. The attacker might carefully craft attacks vectors such that the mode probability of an incorrect mode be large as that of the true mode, for all the time. If this happens, RIDS can detect attacks as long as one sensor is clean, but might not be able to identify the correct attack vectors. Consider the case that the attacker launch mimicry attacks but at least one sensor is clean. If the mode estimator chooses incorrect mode, the actuator attack estimates would be incorrect since corrupted sensor is used as a reference sensor, but RIDS would notice that physical dynamics with incorrect actuator attack estimates are inconsistent with the testing sensor reading (uncorrupted sensor). Thus, RIDS will raise the alarm, although the attack vector estimates remain incorrect. If the mode selector chooses the correct mode, then RIDS estimates correct attack vectors as we explained beforehand.
It is noteworthy that launching mimicry attacks requires more knowledge and computational power for the attacker, because the attacker should consider the influence of attacks on physical dynamics.

\textbf{Noise, fault, and attack}
RIDS models measurement noises and process noises of a robot and estimates data corruptions with tracked noise propagation. Under certain confidence level, RIDS would not raise alarm under the influence of noises. In this paper, data corruptions model the effects of actuator and sensor attacks. In fact, unintentional actuator and sensor fault/malfunctioning may also result in the detection of data corruptions. From security and safety perspective, both fault and attack may thwart mission execution, and RIDS conducts the detection without distinguishing the two cases. Approaches to distinguish faults and attacks can leverage statistical or knowledge-based fault modeling \cite{da2012knowledge, park2015sensor}, which is beyond the scope of this paper. For the attacks that can be detected and identified, RIDS cannot distinguish different attack channels which result in the same attack vectors.

\textbf{Mode set selection}
In the multi-mode estimation engine design, each mode represents a hypothesis that particular reference sensors are clean and the rest of testing sensors are potentially corrupted. The number of modes ${\mathcal M}$ grow linearly with the number of onboard sensors in a robot, and the computational complexity grows accordingly. The choice of ${\mathcal M}$ is a trade-off between computational complexity and detection accuracy. In particular, with $m$ sensing workflows, the number of possible attack conditions grows exponentially where ${\mathcal M_{complete}}=2^m-1$ (exclude the condition when all sensors are corrupted). Noticeably, the mode set that assumes only one reference sensor remains the same detection capability with that which considers ${\mathcal M_{complete}}=2^m-1$ as demonstrated in~\cite{kim2016attack}. Hence, we employ the current mode set in the multi-mode estimation engine in favor of computational complexity. In the NUISE algorithm of each mode, reference sensor readings are leveraged in the estimation process, and their sensing variances are propagated into the the state and attack vector estimates (see the propagation of $R_{2,k}^j$ in Algorithm~\ref{nalgo2_full}). When multiple reference sensors are available in a mode, the estimation process can perform \emph{sensor fusion}~\cite{roecker1988comparison} and reduce estimation variances. Hence, adding more modes into the estimation engine increases the detection accuracy when multiple sensors remain uncorrupted. Table~\ref{tab:sensor_settings} shows actuator attack vector variance comparison in different modes. Noticeably, the mode which assumes all sensors are uncorrupted generates lowest variance.
\begin{table}[!ht]
\caption{Actuator attack vector quantification variance under different mode.}
\label{tab:sensor_settings}
\centering
\begin{tabular}{l|l|l}
\hline
\textbf{Reference Sensor(s)} & \textbf{Var on} $V_l$ ($\times 10^{-5}$) & \textbf{Var on} $V_r$ ($\times 10^{-5}$)\\
\hline
\hline
IPS & 2.39 & 1.94\\
\hline
Wheel encoder & 2.76 & 2.04\\
\hline
LiDAR & 21.7 & 20.3\\
\hline
all sensors & 2.32 & 1.88\\
\hline
\end{tabular}
\end{table}

%%%%%%%%%%%%%%%%%%%%%%%%%%%%%%%%%%%%%%%%%%%%%%%%%%%%%%%%%%%%%%%%%%
% \textbf{Intrusion response}
% By launching sensor attacks and actuator attacks, attackers can influence or potentially take over a robot; e.g., deviating the robot from its planned route, or colliding with neighboring robots. Once attacks are detected, how to ensure a robot can continue to function or minimize its impact is an open question~\cite{cardenas2008secure}. With RIDS, the planner can take immediate response actions to ensure safety of the victim and other robots, e.g. emergency power cut-off, or broadcasting alarms to its neighboring robots, etc. %For instance, RIDS can override the current control loop and initiate emergency control. For multiple-robot systems, RIDS can also broadcast alarms to its neighboring robots, and re-plan paths to avoid collisions. 
% %The current RIDS provides an adaptive response, which allows the robot to complete mission when multiple sensors are under attack. 
% One of our future work is to synthesize and test computationally efficient re-planning algorithms as intrusion response.
%%%%%%%%%%%%%%%%%%%%%%%%%%%%%%%%%%%%%%%%%%%%%%%%%%%%%%%%%%%%%%%%%%

\textbf{Sensor capability} This paper only considers mobile robot states in terms of movements. Hence, sensors that measure other statuses of robots, e.g., temperature, tire pressure, are out of scope. During the estimation process, the NUISE Algorithm estimates robot states using reference sensor readings in each mode. A requirement is that the reference sensor(s) in each mode are capable of reconstructing all robot states in a control iteration. However, some sensors might not be utilized to reconstruct all states of a robot. For instance, consider a ground robot equipped with a magnetometer which can only measure the orientation $\theta_M$ of the robot. Since robot states are described as $(x_k,y_k,\theta_k)$, the measurement from magnetometer cannot reconstruct the robot states. If the robot runs RIDS, then the mode that only takes magnetometer as reference sensor will fail to estimate the states and the attack vectors. Under such cases, RIDS designers can group multiple sensors together to ensure the reference sensors of each mode can reconstruct robot states. For instance, the magnetometer can be grouped together with a GPS sensor and use $h(\textbf{x}_{k})=[x_k,y_k,\theta_k,\theta_k]^T$ as their joint measurement model.

\section{Related Work}
The security of robots and other cyber-physical systems (CPS) has been attracting increasing attention. In this section, we review some preliminary studies concerning several topics related to this work.

\textbf{Existing attacks on robots} Preliminary works identified attacks launched through different channels, including physical damage, network communication, signal interference, malware, etc. Koscher et al. demonstrated that virtually any ECUs inside the internal vehicular network of a modern vehicle can be infiltrated through physical access~\cite{koscher2010experimental}. Checkoway et al. further demonstrated that remote exploitation through wireless channels, such as Bluetooth or cellular radio, is also possible ~\cite{checkoway2011comprehensive}. AnonSec group took over a NASA Global Hawk drone and tried to crash the drone into the ocean by breaking into internal network~\cite{anonsec}. Several studies~\cite{humphreys2008assessing,tippenhauer2011requirements,zaragoza2013spoofing} investigated spoofing attacks targeted on civilian GPS signals. Some researchers have also implemented deceptive spoofers and conducted proof of concept attack experiments~\cite{humphreys2008assessing,montgomery2009receiver}. Son et al.~\cite{son2015rocking} demonstrated that resonant frequency of sound could be used to incapacitate a drone through its gyroscope sensor. Although at an early development stage, robot malware has already debuted. Sasi~\cite{maldrone} developed a backdoor program which allows attackers to control the drone remotely.

\textbf{Intrusion detection for CPSs} 
State estimation theory has been utilized to detect sensor attacks for linear cyber-physical systems in recent works~\cite{bezzo2014attack,mo2010false,park2015sensor,pajic2015attack}. Several works~\cite{fawzi2014secure,yong2015resilient,pasqualetti2013attack} study both actuator and sensor attacks for linear cyber-physical systems with estimation theory. In contrast, most real-world robots are modeled as nonlinear systems, such as Khepera and UAVs.
In~\cite{fawzi2014secure,pajic2015attack,park2015sensor,pasqualetti2013attack}, processing and measurement noises rooted in actuators and sensors are not considered or considered with bounded support. In contrast, real-world robots are subject to stochastic noises with unbounded support. Shoukry et al.~\cite{shoukry2015pycra} proposed a sensor attack detection approach against signal interference attacks by verifying randomly inserted probes. A few studies in sensing systems proposed GNSS attack detection techniques~\cite{montgomery2009receiver,psiaki2011civilian}. Montgomery et al. proposed to detect GNSS attacks by exploiting the effects of intentional high-frequency antenna motion~\cite{montgomery2009receiver}. Psiaki et al. validated the correctness of civilian GPS signals using dual-receiver correlation of military signals~\cite{psiaki2011civilian}. Some of these techniques require homogeneous sensors or extra hardware to enable a comparison between sensors, and some require cryptography for authentication purposes. In robot systems, sensors usually measure different physical signal configurations. Extra hardware brings additional costs and burdens for power supply and weight carrying.

\section{Conclusion and Future Work}
Sensor attacks and actuator attacks targeted on mobile robots impose a huge security threat. In this study, we propose the first practical robot intrusion detection system framework called RIDS, which is capable of detecting, identifying and quantifying both types of attacks. %Leveraging the correlations between sensor readings and control commands, we develop a multi-mode nonlinear unknown input and state estimation algorithm in the detection of both attacks. Using statistical hypothesis testing, RIDS can detect attacks under certain confidence level, identify attack types, and quantify the attack vectors for further analysis and intrusion response. RIDS provides a resilient approach for attack detection in real world mobile robots. RIDS considers unbounded Gaussian noises for both actuators and sensors, and generates detection results with probabilistic confidence. 
We conduct experiments on Khepera testbed which runs a motion planning mission. Our evaluation results show satisfactory detection performance under high significance levels with negligible detection delays. Future work will focus on designing and synthesizing computationally efficient intrusion response algorithms after detection.

% We conduct experiments on Khepera testbed which embodies real world robot settings and mission. Our evaluation results show satisfactory detection performance under high significance levels with negligible detection delays. Attack vector estimates provide precise quantitative information about the attacks. In our experiments, we demonstrate that as long as at least one sensor is not compromised, RIDS could detect both actuator attacks and sensor attacks.

% Future work will focus on conducting more empirical experiments on implementing RIDS in various mobile robots. Moreover, introducing artificial diversity into sensing workflows could be a promising direction to improve the resilience of the RIDS against more sophisticated attacks. Using the categorized and quantified detection information, how to design and synthesize computationally efficient intrusion response algorithms would also be an interesting topic to investigate ahead.
%Although the related work and empirical security analysis indicates that some robot sensors are more vulnerable to attacks, a more trustworthy IDS which does not rely on prior knowledge of trusted and untrusted sensors would be an interesting topic to investigate ahead.

%\bibliographystyle{ACM-Reference-Format}
%\bibliography{biblio}


\begin{thebibliography}{00}

%%% the Web address.

\ifx \showCODEN    \undefined \def \showCODEN     #1{\unskip}     \fi
\ifx \showDOI      \undefined \def \showDOI       #1{#1}\fi
\ifx \showISBNx    \undefined \def \showISBNx     #1{\unskip}     \fi
\ifx \showISBNxiii \undefined \def \showISBNxiii  #1{\unskip}     \fi
\ifx \showISSN     \undefined \def \showISSN      #1{\unskip}     \fi
\ifx \showLCCN     \undefined \def \showLCCN      #1{\unskip}     \fi
\ifx \shownote     \undefined \def \shownote      #1{#1}          \fi
\ifx \showarticletitle \undefined \def \showarticletitle #1{#1}   \fi
\ifx \showURL      \undefined \def \showURL       {\relax}        \fi
% invisible to TeX
\providecommand\bibfield[2]{#2}
\providecommand\bibinfo[2]{#2}
\providecommand\natexlab[1]{#1}
\providecommand\showeprint[2][]{arXiv:#2}

\bibitem[\protect\citeauthoryear{??}{rob}{2009}]%
        {robotics_stat1}
 \bibinfo{year}{2009}\natexlab{}.
\newblock \bibinfo{title}{Executive Summary of World Robotics 2009}.
\newblock
  \bibinfo{howpublished}{\url{http://www.dis.uniroma1.it/\~deluca/rob1_en/2009_WorldRobotics_ExecSummary.pdf}}.
    (\bibinfo{year}{2009}).
\newblock


\bibitem[\protect\citeauthoryear{??}{tes}{2016}]%
        {teslahack}
 \bibinfo{year}{2016}\natexlab{}.
\newblock \bibinfo{title}{Car Hacking Research: Remote Attack Tesla Motors.
  {K}een {S}ecurity {L}ab of {T}encent}.
\newblock
  \bibinfo{howpublished}{\url{http://keenlab.tencent.com/en/2016/09/19/Keen-Security-Lab-of-Tencent-Car-Hacking-Research-Remote-Attack-to-Tesla-Cars/}}.
    (\bibinfo{year}{2016}).
\newblock


\bibitem[\protect\citeauthoryear{??}{khe}{2016}]%
        {khepera}
 \bibinfo{year}{2016}\natexlab{}.
\newblock \bibinfo{title}{K-Team Mobile Robotics - {K}hepera {III}}.
\newblock
  \bibinfo{howpublished}{\url{http://www.k-team.com/mobile-robotics-products/old-products/khepera-iii}}.
    (\bibinfo{year}{2016}).
\newblock


\bibitem[\protect\citeauthoryear{??}{kor}{2016}]%
        {korebot}
 \bibinfo{year}{2016}\natexlab{}.
\newblock \bibinfo{title}{K-Team Mobile Robotics - {K}oreBot {II}}.
\newblock
  \bibinfo{howpublished}{\url{http://www.k-team.com/mobile-robotics-products/old-products/korebot-ii}}.
    (\bibinfo{year}{2016}).
\newblock


\bibitem[\protect\citeauthoryear{Akdemir, Karakoyunlu, Padir, and
  Sunar}{Akdemir et~al\mbox{.}}{2010}]%
        {akdemir2010emerging}
\bibfield{author}{\bibinfo{person}{Kahraman~D Akdemir}, \bibinfo{person}{Deniz
  Karakoyunlu}, \bibinfo{person}{Taskin Padir}, {and} \bibinfo{person}{Berk
  Sunar}.} \bibinfo{year}{2010}\natexlab{}.
\newblock \showarticletitle{An emerging threat: eve meets a robot}.
\newblock In \bibinfo{booktitle}{{\em Trusted Systems}}.
\newblock


\bibitem[\protect\citeauthoryear{Bavdekar, Deshpande, and Patwardhan}{Bavdekar
  et~al\mbox{.}}{2011}]%
        {bavdekar2011identification}
\bibfield{author}{\bibinfo{person}{Vinay~A Bavdekar}, \bibinfo{person}{Anjali~P
  Deshpande}, {and} \bibinfo{person}{Sachin~C Patwardhan}.}
  \bibinfo{year}{2011}\natexlab{}.
\newblock \showarticletitle{Identification of process and measurement noise
  covariance for state and parameter estimation using extended Kalman filter}.
\newblock \bibinfo{journal}{{\em Journal of Process control\/}}
  (\bibinfo{year}{2011}).
\newblock


\bibitem[\protect\citeauthoryear{Bezzo, Weimer, Pajic, Sokolsky, Pappas, and
  Lee}{Bezzo et~al\mbox{.}}{2014}]%
        {bezzo2014attack}
\bibfield{author}{\bibinfo{person}{Nicola Bezzo}, \bibinfo{person}{James
  Weimer}, \bibinfo{person}{Miroslav Pajic}, \bibinfo{person}{Oleg Sokolsky},
  \bibinfo{person}{George~J Pappas}, {and} \bibinfo{person}{Insup Lee}.}
  \bibinfo{year}{2014}\natexlab{}.
\newblock \showarticletitle{Attack resilient state estimation for autonomous
  robotic systems}. In \bibinfo{booktitle}{{\em IROS}}.
\newblock


\bibitem[\protect\citeauthoryear{Castro, Liskov, et~al\mbox{.}}{Castro
  et~al\mbox{.}}{1999}]%
        {castro1999practical}
\bibfield{author}{\bibinfo{person}{Miguel Castro}, \bibinfo{person}{Barbara
  Liskov}, {et~al\mbox{.}}} \bibinfo{year}{1999}\natexlab{}.
\newblock \showarticletitle{Practical Byzantine fault tolerance}. In
  \bibinfo{booktitle}{{\em OSDI}}.
\newblock


\bibitem[\protect\citeauthoryear{Checkoway, McCoy, Kantor, Anderson, Shacham,
  Savage, Koscher, Czeskis, Roesner, Kohno, et~al\mbox{.}}{Checkoway
  et~al\mbox{.}}{2011}]%
        {checkoway2011comprehensive}
\bibfield{author}{\bibinfo{person}{Stephen Checkoway}, \bibinfo{person}{Damon
  McCoy}, \bibinfo{person}{Brian Kantor}, \bibinfo{person}{Danny Anderson},
  \bibinfo{person}{Hovav Shacham}, \bibinfo{person}{Stefan Savage},
  \bibinfo{person}{Karl Koscher}, \bibinfo{person}{Alexei Czeskis},
  \bibinfo{person}{Franziska Roesner}, \bibinfo{person}{Tadayoshi Kohno},
  {et~al\mbox{.}}} \bibinfo{year}{2011}\natexlab{}.
\newblock \showarticletitle{Comprehensive Experimental Analyses of Automotive
  Attack Surfaces.}. In \bibinfo{booktitle}{{\em USENIX Security}}.
\newblock


\bibitem[\protect\citeauthoryear{Chen, Patton, and Zhang}{Chen
  et~al\mbox{.}}{1996}]%
        {chen1996design}
\bibfield{author}{\bibinfo{person}{Jie Chen}, \bibinfo{person}{Ron~J Patton},
  {and} \bibinfo{person}{Hong-Yue Zhang}.} \bibinfo{year}{1996}\natexlab{}.
\newblock \showarticletitle{Design of unknown input observers and robust fault
  detection filters}.
\newblock \bibinfo{journal}{{\it Internat. J. Control}} (\bibinfo{year}{1996}).
\newblock


\bibitem[\protect\citeauthoryear{Cheng, Ye, Wang, and Zhou}{Cheng
  et~al\mbox{.}}{2009}]%
        {cheng2009unbiased}
\bibfield{author}{\bibinfo{person}{Yue Cheng}, \bibinfo{person}{Hao Ye},
  \bibinfo{person}{Yongqiang Wang}, {and} \bibinfo{person}{Donghua Zhou}.}
  \bibinfo{year}{2009}\natexlab{}.
\newblock \showarticletitle{Unbiased minimum-variance state estimation for
  linear systems with unknown input}.
\newblock \bibinfo{journal}{{\em Automatica\/}} (\bibinfo{year}{2009}).
\newblock


\bibitem[\protect\citeauthoryear{Cho, Chen, and Ding}{Cho
  et~al\mbox{.}}{2004}]%
        {cho2004redundancy}
\bibfield{author}{\bibinfo{person}{Jung~Jin Cho}, \bibinfo{person}{Yong Chen},
  {and} \bibinfo{person}{Yu Ding}.} \bibinfo{year}{2004}\natexlab{}.
\newblock \bibinfo{title}{Redundancy Analysis of Linear Sensor Systems and Its
  Applications}.
\newblock
  \bibinfo{howpublished}{\url{https://ww.orchampion.org/content/download/55235/522615/file/redundancy.pdf}}.
    (\bibinfo{year}{2004}).
\newblock


\bibitem[\protect\citeauthoryear{Chow and Willsky}{Chow and Willsky}{1984}]%
        {chow1984analytical}
\bibfield{author}{\bibinfo{person}{EYEY Chow} {and} \bibinfo{person}{A
  Willsky}.} \bibinfo{year}{1984}\natexlab{}.
\newblock \showarticletitle{Analytical redundancy and the design of robust
  failure detection systems}.
\newblock \bibinfo{journal}{{\em IEEE Transactions on automatic control\/}}
  (\bibinfo{year}{1984}).
\newblock


\bibitem[\protect\citeauthoryear{da~Silva, Saxena, Balaban, and
  Goebel}{da~Silva et~al\mbox{.}}{2012}]%
        {da2012knowledge}
\bibfield{author}{\bibinfo{person}{Jonny~Carlos da Silva},
  \bibinfo{person}{Abhinav Saxena}, \bibinfo{person}{Edward Balaban}, {and}
  \bibinfo{person}{Kai Goebel}.} \bibinfo{year}{2012}\natexlab{}.
\newblock \showarticletitle{A knowledge-based system approach for sensor fault
  modeling, detection and mitigation}.
\newblock \bibinfo{journal}{{\em Expert Systems with Applications\/}}
  (\bibinfo{year}{2012}).
\newblock


\bibitem[\protect\citeauthoryear{Darouach and Zasadzinski}{Darouach and
  Zasadzinski}{1997}]%
        {darouach1997unbiased}
\bibfield{author}{\bibinfo{person}{Mohamed Darouach} {and}
  \bibinfo{person}{Michel Zasadzinski}.} \bibinfo{year}{1997}\natexlab{}.
\newblock \showarticletitle{Unbiased minimum variance estimation for systems
  with unknown exogenous inputs}.
\newblock \bibinfo{journal}{{\em Automatica\/}} (\bibinfo{year}{1997}).
\newblock


\bibitem[\protect\citeauthoryear{de~Almeida, de~Carvalho~Ferreira, and
  Val{\'e}rio}{de~Almeida et~al\mbox{.}}{2013}]%
        {de2013microkernel}
\bibfield{author}{\bibinfo{person}{Rodrigo Maximiano~Antunes de Almeida},
  \bibinfo{person}{Luis~Henrique de Carvalho~Ferreira}, {and}
  \bibinfo{person}{Carlos~Henrique Val{\'e}rio}.}
  \bibinfo{year}{2013}\natexlab{}.
\newblock \showarticletitle{Microkernel development for embedded systems}.
\newblock \bibinfo{journal}{{\em Journal of Software Engineering and
  Applications\/}} (\bibinfo{year}{2013}).
\newblock


\bibitem[\protect\citeauthoryear{De~Nicolao, Sparacino, and Cobelli}{De~Nicolao
  et~al\mbox{.}}{1997}]%
        {de1997nonparametric}
\bibfield{author}{\bibinfo{person}{Giuseppe De~Nicolao},
  \bibinfo{person}{Giovanni Sparacino}, {and} \bibinfo{person}{Claudio
  Cobelli}.} \bibinfo{year}{1997}\natexlab{}.
\newblock \showarticletitle{Nonparametric input estimation in physiological
  systems: problems, methods, and case studies}.
\newblock \bibinfo{journal}{{\em Automatica\/}} (\bibinfo{year}{1997}).
\newblock


\bibitem[\protect\citeauthoryear{Di~Natale}{Di~Natale}{2008}]%
        {di2008understanding}
\bibfield{author}{\bibinfo{person}{Marco Di~Natale}.}
  \bibinfo{year}{2008}\natexlab{}.
\newblock \showarticletitle{Understanding and using the Controller Area
  network}.
\newblock  (\bibinfo{year}{2008}).
\newblock


\bibitem[\protect\citeauthoryear{Durrant-Whyte and Bailey}{Durrant-Whyte and
  Bailey}{2006}]%
        {durrant2006simultaneous}
\bibfield{author}{\bibinfo{person}{Hugh Durrant-Whyte} {and}
  \bibinfo{person}{Tim Bailey}.} \bibinfo{year}{2006}\natexlab{}.
\newblock \showarticletitle{Simultaneous localization and mapping}.
\newblock \bibinfo{journal}{{\em IEEE robotics \& automation magazine\/}}
  (\bibinfo{year}{2006}).
\newblock


\bibitem[\protect\citeauthoryear{Fawzi, Tabuada, and Diggavi}{Fawzi
  et~al\mbox{.}}{2014}]%
        {fawzi2014secure}
\bibfield{author}{\bibinfo{person}{Hamza Fawzi}, \bibinfo{person}{Paulo
  Tabuada}, {and} \bibinfo{person}{Suhas Diggavi}.}
  \bibinfo{year}{2014}\natexlab{}.
\newblock \showarticletitle{Secure estimation and control for cyber-physical
  systems under adversarial attacks}.
\newblock \bibinfo{journal}{{\it IEEE Trans. Automat. Control}}
  (\bibinfo{year}{2014}).
\newblock


\bibitem[\protect\citeauthoryear{Flynn}{Flynn}{1985}]%
        {flynn1985redundant}
\bibfield{author}{\bibinfo{person}{Anita~M Flynn}.}
  \bibinfo{year}{1985}\natexlab{}.
\newblock \showarticletitle{Redundant sensors for mobile robot navigation}.
\newblock  (\bibinfo{year}{1985}).
\newblock


\bibitem[\protect\citeauthoryear{Hou and Patton}{Hou and Patton}{1998}]%
        {hou1998optimal}
\bibfield{author}{\bibinfo{person}{M Hou} {and} \bibinfo{person}{RJ Patton}.}
  \bibinfo{year}{1998}\natexlab{}.
\newblock \showarticletitle{Optimal filtering for systems with unknown inputs}.
\newblock \bibinfo{journal}{{\it IEEE Trans. Automat. Control}}
  (\bibinfo{year}{1998}).
\newblock


\bibitem[\protect\citeauthoryear{Humphreys, Ledvina, Psiaki, O’Hanlon, and
  Kintner~Jr}{Humphreys et~al\mbox{.}}{2008}]%
        {humphreys2008assessing}
\bibfield{author}{\bibinfo{person}{Todd~E Humphreys}, \bibinfo{person}{Brent~M
  Ledvina}, \bibinfo{person}{Mark~L Psiaki}, \bibinfo{person}{Brady~W
  O’Hanlon}, {and} \bibinfo{person}{Paul~M Kintner~Jr}.}
  \bibinfo{year}{2008}\natexlab{}.
\newblock \showarticletitle{Assessing the spoofing threat: Development of a
  portable {GPS} civilian spoofer}. In \bibinfo{booktitle}{{\em ION GNSS}}.
\newblock


\bibitem[\protect\citeauthoryear{IDC}{IDC}{2016}]%
        {idc}
\bibfield{author}{\bibinfo{person}{IDC}.} \bibinfo{year}{2016}\natexlab{}.
\newblock \bibinfo{title}{MANUFACTURING Press Release}.
\newblock
  \bibinfo{howpublished}{\url{http://www.idc.com/getdoc.jsp?containerId=prUS41046916}}.
    (\bibinfo{year}{2016}).
\newblock


\bibitem[\protect\citeauthoryear{Jazwinski}{Jazwinski}{2007}]%
        {jazwinski2007stochastic}
\bibfield{author}{\bibinfo{person}{Andrew~H Jazwinski}.}
  \bibinfo{year}{2007}\natexlab{}.
\newblock \bibinfo{booktitle}{{\em Stochastic processes and filtering theory}}.
\newblock \bibinfo{publisher}{Courier Corporation}.
\newblock


\bibitem[\protect\citeauthoryear{Jetto, Longhi, and Venturini}{Jetto
  et~al\mbox{.}}{1999}]%
        {jetto1999development}
\bibfield{author}{\bibinfo{person}{Leopoldo Jetto}, \bibinfo{person}{Sauro
  Longhi}, {and} \bibinfo{person}{Giuseppe Venturini}.}
  \bibinfo{year}{1999}\natexlab{}.
\newblock \showarticletitle{Development and experimental validation of an
  adaptive extended Kalman filter for the localization of mobile robots}.
\newblock \bibinfo{journal}{{\em Robotics and Automation, IEEE Transactions
  on\/}} (\bibinfo{year}{1999}).
\newblock


\bibitem[\protect\citeauthoryear{Kailath, Sayed, and Hassibi}{Kailath
  et~al\mbox{.}}{2000}]%
        {kailath2000linear}
\bibfield{author}{\bibinfo{person}{Thomas Kailath}, \bibinfo{person}{Ali~H
  Sayed}, {and} \bibinfo{person}{Babak Hassibi}.}
  \bibinfo{year}{2000}\natexlab{}.
\newblock \bibinfo{booktitle}{{\em Linear estimation}}.
\newblock \bibinfo{publisher}{Prentice Hall}.
\newblock


\bibitem[\protect\citeauthoryear{Karaman and Frazzoli}{Karaman and
  Frazzoli}{2011}]%
        {karaman2011sampling}
\bibfield{author}{\bibinfo{person}{Sertac Karaman} {and}
  \bibinfo{person}{Emilio Frazzoli}.} \bibinfo{year}{2011}\natexlab{}.
\newblock \showarticletitle{Sampling-based algorithms for optimal motion
  planning}.
\newblock \bibinfo{journal}{{\em The International Journal of Robotics
  Research\/}} (\bibinfo{year}{2011}).
\newblock


\bibitem[\protect\citeauthoryear{Khandelwal}{Khandelwal}{2015}]%
        {maldrone}
\bibfield{author}{\bibinfo{person}{Swati Khandelwal}.}
  \bibinfo{year}{2015}\natexlab{}.
\newblock \bibinfo{title}{MalDrone - First Ever Backdoor Malware for Drones}.
\newblock
  \bibinfo{howpublished}{\url{http://thehackernews.com/2015/01/MalDrone-backdoor-drone-malware.html}}.
    (\bibinfo{year}{2015}).
\newblock


\bibitem[\protect\citeauthoryear{Kim, Guo, Zhu, and Liu}{Kim
  et~al\mbox{.}}{2017}]%
        {kim2016attack}
\bibfield{author}{\bibinfo{person}{Hunmin Kim}, \bibinfo{person}{Pinyao Guo},
  \bibinfo{person}{Minghui Zhu}, {and} \bibinfo{person}{Peng Liu}.}
  \bibinfo{year}{2017}\natexlab{}.
\newblock \showarticletitle{Attack-resilient Estimation of Switched Nonlinear
  Cyber-Physical Systems, to appear}. In \bibinfo{booktitle}{{\em American
  Control Conference (ACC)}}.
\newblock


\bibitem[\protect\citeauthoryear{Kitanidis}{Kitanidis}{1987}]%
        {kitanidis1987unbiased}
\bibfield{author}{\bibinfo{person}{Peter~K Kitanidis}.}
  \bibinfo{year}{1987}\natexlab{}.
\newblock \showarticletitle{Unbiased minimum-variance linear state estimation}.
\newblock \bibinfo{journal}{{\em Automatica\/}} (\bibinfo{year}{1987}).
\newblock


\bibitem[\protect\citeauthoryear{Koscher, Czeskis, Roesner, Patel, Kohno,
  Checkoway, McCoy, Kantor, Anderson, Shacham, et~al\mbox{.}}{Koscher
  et~al\mbox{.}}{2010}]%
        {koscher2010experimental}
\bibfield{author}{\bibinfo{person}{Karl Koscher}, \bibinfo{person}{Alexei
  Czeskis}, \bibinfo{person}{Franziska Roesner}, \bibinfo{person}{Shwetak
  Patel}, \bibinfo{person}{Tadayoshi Kohno}, \bibinfo{person}{Stephen
  Checkoway}, \bibinfo{person}{Damon McCoy}, \bibinfo{person}{Brian Kantor},
  \bibinfo{person}{Danny Anderson}, \bibinfo{person}{Hovav Shacham},
  {et~al\mbox{.}}} \bibinfo{year}{2010}\natexlab{}.
\newblock \showarticletitle{Experimental security analysis of a modern
  automobile}. In \bibinfo{booktitle}{{\em S\&P}}.
\newblock


\bibitem[\protect\citeauthoryear{Kotecha and Djuric}{Kotecha and
  Djuric}{2003}]%
        {kotecha2003gaussian}
\bibfield{author}{\bibinfo{person}{Jayesh~H Kotecha} {and}
  \bibinfo{person}{Petar~M Djuric}.} \bibinfo{year}{2003}\natexlab{}.
\newblock \showarticletitle{Gaussian particle filtering}.
\newblock \bibinfo{journal}{{\em IEEE Transactions on signal processing\/}}
  (\bibinfo{year}{2003}).
\newblock


\bibitem[\protect\citeauthoryear{Krishnamachari and Iyengar}{Krishnamachari and
  Iyengar}{2004}]%
        {krishnamachari2004distributed}
\bibfield{author}{\bibinfo{person}{Bhaskar Krishnamachari} {and}
  \bibinfo{person}{Sitharama Iyengar}.} \bibinfo{year}{2004}\natexlab{}.
\newblock \showarticletitle{Distributed Bayesian algorithms for fault-tolerant
  event region detection in wireless sensor networks}.
\newblock \bibinfo{journal}{{\it IEEE Trans. Comput.}} (\bibinfo{year}{2004}).
\newblock


\bibitem[\protect\citeauthoryear{Kumar and Michael}{Kumar and Michael}{2012}]%
        {kumar2012opportunities}
\bibfield{author}{\bibinfo{person}{Vijay Kumar} {and} \bibinfo{person}{Nathan
  Michael}.} \bibinfo{year}{2012}\natexlab{}.
\newblock \showarticletitle{Opportunities and challenges with autonomous micro
  aerial vehicles}.
\newblock \bibinfo{journal}{{\em IJRR\/}} (\bibinfo{year}{2012}).
\newblock


\bibitem[\protect\citeauthoryear{Leonard, How, Teller, Berger, Campbell, Fiore,
  Fletcher, Frazzoli, Huang, Karaman, et~al\mbox{.}}{Leonard
  et~al\mbox{.}}{2008}]%
        {leonard2008perception}
\bibfield{author}{\bibinfo{person}{John Leonard}, \bibinfo{person}{Jonathan
  How}, \bibinfo{person}{Seth Teller}, \bibinfo{person}{Mitch Berger},
  \bibinfo{person}{Stefan Campbell}, \bibinfo{person}{Gaston Fiore},
  \bibinfo{person}{Luke Fletcher}, \bibinfo{person}{Emilio Frazzoli},
  \bibinfo{person}{Albert Huang}, \bibinfo{person}{Sertac Karaman},
  {et~al\mbox{.}}} \bibinfo{year}{2008}\natexlab{}.
\newblock \showarticletitle{A perception-driven autonomous urban vehicle}.
\newblock \bibinfo{journal}{{\em Journal of Field Robotics\/}}
  (\bibinfo{year}{2008}).
\newblock


\bibitem[\protect\citeauthoryear{Li, Srinivasan, and Wu}{Li
  et~al\mbox{.}}{2008}]%
        {li2008pvfs}
\bibfield{author}{\bibinfo{person}{Feng Li}, \bibinfo{person}{Avinash
  Srinivasan}, {and} \bibinfo{person}{Jie Wu}.}
  \bibinfo{year}{2008}\natexlab{}.
\newblock \showarticletitle{PVFS: a probabilistic voting-based filtering scheme
  in wireless sensor networks}.
\newblock \bibinfo{journal}{{\em International Journal of Security and
  Networks\/}} (\bibinfo{year}{2008}).
\newblock


\bibitem[\protect\citeauthoryear{Litman}{Litman}{2014}]%
        {autotrend}
\bibfield{author}{\bibinfo{person}{Todd Litman}.}
  \bibinfo{year}{2014}\natexlab{}.
\newblock \bibinfo{title}{Autonomous vehicle implementation predictions.
  {I}mplications for transport planning.}
\newblock \bibinfo{howpublished}{\url{http://www.vtpi.org/avip.pdf}}.
  (\bibinfo{year}{2014}).
\newblock


\bibitem[\protect\citeauthoryear{Liu and Hwang}{Liu and Hwang}{2011}]%
        {liu2011robust}
\bibfield{author}{\bibinfo{person}{W Liu} {and} \bibinfo{person}{I Hwang}.}
  \bibinfo{year}{2011}\natexlab{}.
\newblock \showarticletitle{Robust estimation and fault detection and isolation
  algorithms for stochastic linear hybrid systems with unknown fault input}.
\newblock \bibinfo{journal}{{\em IET control theory \& applications\/}}
  (\bibinfo{year}{2011}).
\newblock


\bibitem[\protect\citeauthoryear{Miller and Valasek}{Miller and
  Valasek}{2015}]%
        {miller2015remote}
\bibfield{author}{\bibinfo{person}{Charlie Miller} {and} \bibinfo{person}{Chris
  Valasek}.} \bibinfo{year}{2015}\natexlab{}.
\newblock \showarticletitle{Remote exploitation of an unaltered passenger
  vehicle}.
\newblock \bibinfo{journal}{{\em Black Hat USA\/}} (\bibinfo{year}{2015}).
\newblock


\bibitem[\protect\citeauthoryear{Mo, Garone, Casavola, and Sinopoli}{Mo
  et~al\mbox{.}}{2010}]%
        {mo2010false}
\bibfield{author}{\bibinfo{person}{Yilin Mo}, \bibinfo{person}{Emanuele
  Garone}, \bibinfo{person}{Alessandro Casavola}, {and} \bibinfo{person}{Bruno
  Sinopoli}.} \bibinfo{year}{2010}\natexlab{}.
\newblock \showarticletitle{False data injection attacks against state
  estimation in wireless sensor networks}. In \bibinfo{booktitle}{{\em CDC}}.
\newblock


\bibitem[\protect\citeauthoryear{Montgomery, Humphreys, and Ledvina}{Montgomery
  et~al\mbox{.}}{2009}]%
        {montgomery2009receiver}
\bibfield{author}{\bibinfo{person}{Paul~Y Montgomery}, \bibinfo{person}{Todd~E
  Humphreys}, {and} \bibinfo{person}{Brent~M Ledvina}.}
  \bibinfo{year}{2009}\natexlab{}.
\newblock \showarticletitle{Receiver-autonomous spoofing detection:
  Experimental results of a multi-antenna receiver defense against a portable
  civil {GPS} spoofer}. In \bibinfo{booktitle}{{\em ITM}}.
\newblock


\bibitem[\protect\citeauthoryear{Nemra and Aouf}{Nemra and Aouf}{2009}]%
        {nemra2009robust}
\bibfield{author}{\bibinfo{person}{Abdelkrim Nemra} {and}
  \bibinfo{person}{Nabil Aouf}.} \bibinfo{year}{2009}\natexlab{}.
\newblock \showarticletitle{Robust {INS/GPS} sensor fusion for {UAV}
  localization using {SDRE} nonlinear filtering}.
\newblock \bibinfo{journal}{{\em IEEE Sensors Journal\/}}
  (\bibinfo{year}{2009}).
\newblock


\bibitem[\protect\citeauthoryear{Pajic, Tabuada, Lee, and Pappas}{Pajic
  et~al\mbox{.}}{2015}]%
        {pajic2015attack}
\bibfield{author}{\bibinfo{person}{Miroslav Pajic}, \bibinfo{person}{Paulo
  Tabuada}, \bibinfo{person}{Insup Lee}, {and} \bibinfo{person}{George~J
  Pappas}.} \bibinfo{year}{2015}\natexlab{}.
\newblock \showarticletitle{Attack-resilient state estimation in the presence
  of noise}. In \bibinfo{booktitle}{{\em 2015 54th IEEE Conference on Decision
  and Control (CDC)}}.
\newblock


\bibitem[\protect\citeauthoryear{Park, Ivanov, Weimer, Pajic, and Lee}{Park
  et~al\mbox{.}}{2015}]%
        {park2015sensor}
\bibfield{author}{\bibinfo{person}{Junkil Park}, \bibinfo{person}{Radoslav
  Ivanov}, \bibinfo{person}{James Weimer}, \bibinfo{person}{Miroslav Pajic},
  {and} \bibinfo{person}{Insup Lee}.} \bibinfo{year}{2015}\natexlab{}.
\newblock \showarticletitle{Sensor attack detection in the presence of
  transient faults}. In \bibinfo{booktitle}{{\em ICCPS}}.
\newblock


\bibitem[\protect\citeauthoryear{Pasqualetti, Dorfler, and Bullo}{Pasqualetti
  et~al\mbox{.}}{2013}]%
        {pasqualetti2013attack}
\bibfield{author}{\bibinfo{person}{Fabio Pasqualetti}, \bibinfo{person}{Florian
  Dorfler}, {and} \bibinfo{person}{Francesco Bullo}.}
  \bibinfo{year}{2013}\natexlab{}.
\newblock \showarticletitle{Attack detection and identification in
  cyber-physical systems}.
\newblock \bibinfo{journal}{{\em Automatic Control, IEEE Transactions on\/}}
  (\bibinfo{year}{2013}).
\newblock


\bibitem[\protect\citeauthoryear{Paxson}{Paxson}{1999}]%
        {paxson1999bro}
\bibfield{author}{\bibinfo{person}{Vern Paxson}.}
  \bibinfo{year}{1999}\natexlab{}.
\newblock \showarticletitle{Bro: a system for detecting network intruders in
  real-time}.
\newblock \bibinfo{journal}{{\em Computer networks\/}} (\bibinfo{year}{1999}).
\newblock


\bibitem[\protect\citeauthoryear{Petit, Stottelaar, Feiri, and Kargl}{Petit
  et~al\mbox{.}}{2015a}]%
        {2015blackhat}
\bibfield{author}{\bibinfo{person}{Jonathan Petit}, \bibinfo{person}{Bas
  Stottelaar}, \bibinfo{person}{Michael Feiri}, {and} \bibinfo{person}{Frank
  Kargl}.} \bibinfo{year}{2015}\natexlab{a}.
\newblock \showarticletitle{Remote Attacks on Automated Vehicles Sensors:
  Experiments on Camera and LiDAR}. In \bibinfo{booktitle}{{\em Black Hat
  Europe}}.
\newblock
\showURL{%
\url{https://www.blackhat.com/docs/eu-15/materials/eu-15-Petit-Self-Driving-And-Connected-Cars-Fooling-Sensors-And-Tracking-Drivers-wp1.pdf}}


\bibitem[\protect\citeauthoryear{Petit, Stottelaar, Feiri, and Kargl}{Petit
  et~al\mbox{.}}{2015b}]%
        {petit2015remote}
\bibfield{author}{\bibinfo{person}{Jonathan Petit}, \bibinfo{person}{B
  Stottelaar}, \bibinfo{person}{M Feiri}, {and} \bibinfo{person}{F Kargl}.}
  \bibinfo{year}{2015}\natexlab{b}.
\newblock \showarticletitle{Remote attacks on automated vehicles sensors:
  Experiments on camera and lidar}.
\newblock \bibinfo{journal}{{\em Black Hat Europe\/}} (\bibinfo{year}{2015}).
\newblock


\bibitem[\protect\citeauthoryear{Psiaki, O'Hanlon, Bhatti, Shepard, and
  Humphreys}{Psiaki et~al\mbox{.}}{2011}]%
        {psiaki2011civilian}
\bibfield{author}{\bibinfo{person}{Mark~L Psiaki}, \bibinfo{person}{Brady~W
  O'Hanlon}, \bibinfo{person}{Jahshan~A Bhatti}, \bibinfo{person}{Daniel~P
  Shepard}, {and} \bibinfo{person}{Todd~E Humphreys}.}
  \bibinfo{year}{2011}\natexlab{}.
\newblock \showarticletitle{Civilian GPS spoofing detection based on
  dual-receiver correlation of military signals}.
\newblock \bibinfo{journal}{{\em ION GNSS\/}} (\bibinfo{year}{2011}).
\newblock


\bibitem[\protect\citeauthoryear{Rivera, Morari, and Skogestad}{Rivera
  et~al\mbox{.}}{1986}]%
        {rivera1986internal}
\bibfield{author}{\bibinfo{person}{Daniel~E Rivera}, \bibinfo{person}{Manfred
  Morari}, {and} \bibinfo{person}{Sigurd Skogestad}.}
  \bibinfo{year}{1986}\natexlab{}.
\newblock \showarticletitle{Internal model control: PID controller design}.
\newblock \bibinfo{journal}{{\em Industrial \& engineering chemistry process
  design and development\/}} (\bibinfo{year}{1986}).
\newblock


\bibitem[\protect\citeauthoryear{Roecker and McGillem}{Roecker and
  McGillem}{1988}]%
        {roecker1988comparison}
\bibfield{author}{\bibinfo{person}{JA Roecker} {and} \bibinfo{person}{CD
  McGillem}.} \bibinfo{year}{1988}\natexlab{}.
\newblock \showarticletitle{Comparison of two-sensor tracking methods based on
  state vector fusion and measurement fusion}.
\newblock \bibinfo{journal}{{\it IEEE Trans. Aerospace Electron. Systems}}
  (\bibinfo{year}{1988}).
\newblock


\bibitem[\protect\citeauthoryear{Roesch et~al\mbox{.}}{Roesch
  et~al\mbox{.}}{1999}]%
        {roesch1999snort}
\bibfield{author}{\bibinfo{person}{Martin Roesch} {et~al\mbox{.}}}
  \bibinfo{year}{1999}\natexlab{}.
\newblock \showarticletitle{Snort: Lightweight Intrusion Detection for
  Networks.}. In \bibinfo{booktitle}{{\em LISA}}.
\newblock


\bibitem[\protect\citeauthoryear{Russon}{Russon}{2016}]%
        {anonsec}
\bibfield{author}{\bibinfo{person}{Mary~Ann Russon}.}
  \bibinfo{year}{2016}\natexlab{}.
\newblock \bibinfo{title}{{NASA} hack: AnonSec attempts to crash \$222m drone,
  releases secret flight videos and employee data}.
\newblock   (\bibinfo{year}{2016}).
\newblock
\showURL{%
\url{http://www.ibtimes.co.uk/nasa-hack-anonsec-attempts-crash-222m-drone-releases-secret-flight-videos-employee-data-1541254}}


\bibitem[\protect\citeauthoryear{Shoukry, Martin, Yona, Diggavi, and
  Srivastava}{Shoukry et~al\mbox{.}}{2015}]%
        {shoukry2015pycra}
\bibfield{author}{\bibinfo{person}{Yasser Shoukry}, \bibinfo{person}{Paul
  Martin}, \bibinfo{person}{Yair Yona}, \bibinfo{person}{Suhas Diggavi}, {and}
  \bibinfo{person}{Mani Srivastava}.} \bibinfo{year}{2015}\natexlab{}.
\newblock \showarticletitle{PyCRA: Physical Challenge-Response Authentication
  For Active Sensors Under Spoofing Attacks}. In \bibinfo{booktitle}{{\em
  Proceedings of the 22nd ACM SIGSAC Conference on Computer and Communications
  Security}}.
\newblock


\bibitem[\protect\citeauthoryear{Son, Shin, Kim, Park, Noh, Choi, Choi, and
  Kim}{Son et~al\mbox{.}}{2015}]%
        {son2015rocking}
\bibfield{author}{\bibinfo{person}{Yunmok Son}, \bibinfo{person}{Hocheol Shin},
  \bibinfo{person}{Dongkwan Kim}, \bibinfo{person}{Youngseok Park},
  \bibinfo{person}{Juhwan Noh}, \bibinfo{person}{Kibum Choi},
  \bibinfo{person}{Jungwoo Choi}, {and} \bibinfo{person}{Yongdae Kim}.}
  \bibinfo{year}{2015}\natexlab{}.
\newblock \showarticletitle{Rocking drones with intentional sound noise on
  gyroscopic sensors}. In \bibinfo{booktitle}{{\em USENIX Security}}.
\newblock


\bibitem[\protect\citeauthoryear{Tippenhauer, P{\"o}pper, Rasmussen, and
  Capkun}{Tippenhauer et~al\mbox{.}}{2011}]%
        {tippenhauer2011requirements}
\bibfield{author}{\bibinfo{person}{Nils~Ole Tippenhauer},
  \bibinfo{person}{Christina P{\"o}pper}, \bibinfo{person}{Kasper~Bonne
  Rasmussen}, {and} \bibinfo{person}{Srdjan Capkun}.}
  \bibinfo{year}{2011}\natexlab{}.
\newblock \showarticletitle{On the requirements for successful GPS spoofing
  attacks}. In \bibinfo{booktitle}{{\em CCS}}.
\newblock


\bibitem[\protect\citeauthoryear{Urmson, Bagnell, Baker, Hebert, Kelly,
  Rajkumar, Rybski, Scherer, Simmons, Singh, et~al\mbox{.}}{Urmson
  et~al\mbox{.}}{2007}]%
        {urmson2007tartan}
\bibfield{author}{\bibinfo{person}{Chris Urmson}, \bibinfo{person}{J~Andrew
  Bagnell}, \bibinfo{person}{Christopher~R Baker}, \bibinfo{person}{Martial
  Hebert}, \bibinfo{person}{Alonzo Kelly}, \bibinfo{person}{Raj Rajkumar},
  \bibinfo{person}{Paul~E Rybski}, \bibinfo{person}{Sebastian Scherer},
  \bibinfo{person}{Reid Simmons}, \bibinfo{person}{Sanjiv Singh},
  {et~al\mbox{.}}} \bibinfo{year}{2007}\natexlab{}.
\newblock \showarticletitle{Tartan racing: A multi-modal approach to the darpa
  urban challenge}.
\newblock  (\bibinfo{year}{2007}).
\newblock


\bibitem[\protect\citeauthoryear{van Schaik and Heiser}{van Schaik and
  Heiser}{2007}]%
        {van2007high}
\bibfield{author}{\bibinfo{person}{Carl van Schaik} {and}
  \bibinfo{person}{Gernot Heiser}.} \bibinfo{year}{2007}\natexlab{}.
\newblock \showarticletitle{High-performance microkernels and virtualisation on
  ARM and segmented architectures}. In \bibinfo{booktitle}{{\em Proceedings of
  the 1st International Workshop on Microkernels for Embedded Systems, Sydney,
  Australia}}.
\newblock


\bibitem[\protect\citeauthoryear{Warrender, Forrest, and Pearlmutter}{Warrender
  et~al\mbox{.}}{1999}]%
        {warrender1999detecting}
\bibfield{author}{\bibinfo{person}{Christina Warrender},
  \bibinfo{person}{Stephanie Forrest}, {and} \bibinfo{person}{Barak
  Pearlmutter}.} \bibinfo{year}{1999}\natexlab{}.
\newblock \showarticletitle{Detecting intrusions using system calls:
  Alternative data models}. In \bibinfo{booktitle}{{\em Security and Privacy,
  1999. Proceedings of the 1999 IEEE Symposium on}}.
\newblock


\bibitem[\protect\citeauthoryear{Wikipedia}{Wikipedia}{2016}]%
        {wiki:iran}
\bibfield{author}{\bibinfo{person}{Wikipedia}.}
  \bibinfo{year}{2016}\natexlab{}.
\newblock \bibinfo{title}{Iran-{U.S.} {RQ}-170 incident --- Wikipedia{,} The
  Free Encyclopedia}.
\newblock   (\bibinfo{year}{2016}).
\newblock


\bibitem[\protect\citeauthoryear{Yan, Wenyuan, and Liu}{Yan
  et~al\mbox{.}}{2016}]%
        {yan2016can}
\bibfield{author}{\bibinfo{person}{Chen Yan}, \bibinfo{person}{X Wenyuan},
  {and} \bibinfo{person}{Jianhao Liu}.} \bibinfo{year}{2016}\natexlab{}.
\newblock \showarticletitle{Can you trust autonomous vehicles: Contactless
  attacks against sensors of self-driving vehicle}.
\newblock \bibinfo{journal}{{\em DEF CON\/}} (\bibinfo{year}{2016}).
\newblock


\bibitem[\protect\citeauthoryear{Yeung and Ding}{Yeung and Ding}{2003}]%
        {yeung2003host}
\bibfield{author}{\bibinfo{person}{Dit-Yan Yeung} {and} \bibinfo{person}{Yuxin
  Ding}.} \bibinfo{year}{2003}\natexlab{}.
\newblock \showarticletitle{Host-based intrusion detection using dynamic and
  static behavioral models}.
\newblock \bibinfo{journal}{{\em Pattern recognition\/}}
  (\bibinfo{year}{2003}).
\newblock


\bibitem[\protect\citeauthoryear{Yong, Zhu, and Frazzoli}{Yong
  et~al\mbox{.}}{2015}]%
        {yong2015resilient}
\bibfield{author}{\bibinfo{person}{S Yong}, \bibinfo{person}{M Zhu}, {and}
  \bibinfo{person}{E Frazzoli}.} \bibinfo{year}{2015}\natexlab{}.
\newblock \showarticletitle{Resilient state estimation against switching
  attacks on stochastic cyber-physical systems}. In \bibinfo{booktitle}{{\em
  CDC}}.
\newblock


\bibitem[\protect\citeauthoryear{Yong, Zhu, and Frazzoli}{Yong
  et~al\mbox{.}}{2016a}]%
        {SY-MZ-EF:Automatica16}
\bibfield{author}{\bibinfo{person}{S~Z Yong}, \bibinfo{person}{M Zhu}, {and}
  \bibinfo{person}{E Frazzoli}.} \bibinfo{year}{2016}\natexlab{a}.
\newblock \showarticletitle{A unified filter for simultaneous input and state
  estimation of linear discrete-time stochastic systems}.
\newblock \bibinfo{journal}{{\em Automatica\/}} (\bibinfo{year}{2016}).
\newblock


\bibitem[\protect\citeauthoryear{Yong, Zhu, and Frazzoli}{Yong
  et~al\mbox{.}}{2016b}]%
        {yong2016simultaneous}
\bibfield{author}{\bibinfo{person}{Zheng Yong}, \bibinfo{person}{Minghui Zhu},
  {and} \bibinfo{person}{Emilio Frazzoli}.} \bibinfo{year}{2016}\natexlab{b}.
\newblock \showarticletitle{Simultaneous Mode, Input and State Estimation for
  Switched Linear Stochastic Systems}.
\newblock \bibinfo{journal}{{\em arXiv preprint arXiv:1606.08323\/}}
  (\bibinfo{year}{2016}).
\newblock


\bibitem[\protect\citeauthoryear{Zaragoza and Zumalt}{Zaragoza and
  Zumalt}{2013}]%
        {zaragoza2013spoofing}
\bibfield{author}{\bibinfo{person}{S Zaragoza} {and} \bibinfo{person}{E
  Zumalt}.} \bibinfo{year}{2013}\natexlab{}.
\newblock \showarticletitle{Spoofing a Superyacht at Sea}.
\newblock \bibinfo{journal}{{\em Cockrell School of Engineering, UT Austin\/}}
  (\bibinfo{year}{2013}).
\newblock


\end{thebibliography}


\appendix
\section{Appendix}
\subsection{Complete RIDS Design Algorithm}
\begin{algorithm} \caption{Complete Robot Intrusion Detection System (RIDS)} \label{nalgo1_full}
\begin{algorithmic}[1]
\Require %Sensor readings $\textbf{z}_k$ from all sensors; control commands $\textbf{u}_{k-1}$ from control module; 
Initial state estimates $\hat{\textbf{x}}_{0|0}$; robot kinematic function $f(\cdot)$; measurement function $h(\cdot)$
\Ensure Detection decision; attack vector estimates
%\State Initialize: $\hat{\textbf{x}}_{0|0}={\mathbb E}[\textbf{x}_0]$; $P_{0}^x=\textsc{P}_0^x$; $P_{-1}^d = \textsc{P}_{-1}^d$
\State Set parameters $w_s,w_a,c_s,c_a,\alpha_s, \alpha_a$;
\State Initialize;%: $\hat{x}_{0|0}={\mathbb E}[x_0]$; $P_0^x={\mathbb E}[P_0]$; $\mu_k^j=\frac{1}{\mathcal M}$;
% $\tilde{R}_{1,k}=C_{1,0}P_{0}^x C_{1,0}^T + R_{1,0}$;
% $M_{1,0} =(H_{0}^T \tilde{R}_{1,0}^{-1}H_{0})^{-1}H_{0}^T\tilde{R}_{1,0}^{-1}$;
% $P_{0}^{d_1}=(H_{0}^T\tilde{R}_{1,0}^{-1}H_{0})^{-1}$;
% $P_{0}^{xd_1}=(P_{0}^{d_1x})^T=P_{0}^x C_{1,0}^TM_{1,0}^T$; Choose small constant $\epsilon$;
\For{control iteration $k \leftarrow 1$ to $\infty$}
\State Receive control commands $\textbf{u}_{k-1}$;
\State Receive sensor readings $\textbf{z}_k$;
\For{mode $j = 1$ to ${\mathcal M}$}
\State Run NUISE with input ($\textbf{u}_{k-1}$, $\hat{\textbf{x}}_{k-1|k-1}$, $\textbf{z}_{1,k}^j$, $\textbf{z}_{2,k}^j$, $P_{k-1}^x$), 
and generate ($\hat{\textbf{x}}_{k|k}^j$, $\hat{\textbf{d}}_{k}^{s, j}$, $\hat{\textbf{d}}_{k-1}^{a, j}$, $P_{k}^{x,j}$, $P_{k}^{s,j}$, $P_{k-1}^{a,j}$,  ${\mathcal N}_k^j$);
\State $\mu_k^j \leftarrow \max\{{\mathcal N}_k^j \mu_{k-1}^j,\epsilon\}$;
\EndFor
\For{mode $j=1$ to ${\mathcal M}$}
\State $\bar{\mu}_k^j \leftarrow \frac{\mu_k^j}{\sum_{i=1}^{\mathcal M}\mu_k^i}$;
\EndFor
\State Sensor mode accept $J_k \leftarrow {\rm argmax}_j \bar{\mu}_k^j$;
\State Obtain estimates and covariance matrices from $J_k$: $\hat{\textbf{x}}_{k|k}\leftarrow\hat{\textbf{x}}_{k|k}^{J_k}$,
$\hat{\textbf{d}}_{k}^{s}\leftarrow\hat{\textbf{d}}_{k}^{s, J_k}$,
$\hat{\textbf{d}}_{k-1}^{a}\leftarrow\hat{\textbf{d}}_{k-1}^{a, J_k}$,
$P_{k}^x \leftarrow P_k^{x,J_k}$;
\State $b_{k}^s \leftarrow (\hat{{\textbf{d}}_{k}^{s}}^T (P_{k}^{s,J_k})^{-1}\hat{{\textbf{d}}_{k}^{s}}  >\chi_{p=|\hat{\textbf{d}}_{k}^s|}^2(\alpha_s))$;%\footnotemark
\State $b_{k}^a \leftarrow (\hat{\textbf{d}}^{a\,T}_{k-1} (P_{k-1}^{a,J_k})^{-1}\hat{\textbf{d}}^{a}_{k-1}>\chi_{p=|\hat{\textbf{d}}_{k-1}^a|}^2(\alpha_a))$;
\If{$b_{k}^s =True\; \textbf{and} \;\sum_{i = 0}^{w_s-1}b_{k-i}^s \geq c_s$}
\For{each testing sensor $t$ in mode $J_k$}
\State Sensor attack vector estimate for testing sensor $t$: $\hat{\textbf{d}}_{k,t}^s=\sum_{i = 0}^{w_s-1} \hat{\textbf{d}}_{k-i,t}^s / {w_s}$;
\If{$\hat{\textbf{d}}_{k,t}^{s\,T} (P_{k, t}^{s,J_k})^{-1}\hat{\textbf{d}}_{k,t}^{s} \geq \chi_{p=|\hat{\textbf{d}}_{k,t}^{s}|}^2$}
\State Confirm sensor attack on sensor $t$;
\EndIf
\EndFor
% \State Raise alarm for sensor attack on confirmed sensors;
\EndIf
\If {$b_{k}^{a} =True \; \textbf{and} \; \sum_{i = 0}^{w_a-1}b_{k-i}^a \geq c_a$}
     	\State Confirm actuator attack;
        \For{each actuator $t$ in all}
        \State Actuator attack vector estimate for actuator $t$: $\hat{\textbf{d}}_{k-1,t}^a={\sum_{i = 0}^{w_a-1} \hat{\textbf{d}}_{k-1-i,t}^a} / {w_a}$;
        \EndFor
\EndIf\\
%{\footnotemark[\ref{note_algo}]};
%$\|(P_k+P_k^P)^{-\frac{1}{2}}(\textbf{z}_k^{P}-\hat{\textbf{x}}_k)\|_{\infty}>2.58$, then actuator attack exists\footnotemark;
%\State $\|(P_{k-1}^d)^{-\frac{1}{2}}\hat{\textbf{d}}_{k-1}\|_{\infty}>2.58$, then sensor attack exists\footnotemark[\ref{note_algo}];
% \State If $\mathcal{H}_0 : a_k^P = 0$ is rejected{\footnotemark}, then actuator attack exists;
% \State If $\mathcal{H}_0 : d_{k-1} = 0$ is rejected\footnotemark[\ref{note_algo}], then sensor attack exists;
\hspace*{5.2mm}\Return Confirmed attack type(s) and attack target(s); sensor attack vector estimates $\hat{\textbf{d}}_{k,t}^s$ ($t\in\{\textrm{testing sensors in mode }J_k\}$); actuator attack vector estimates $\hat{\textbf{d}}_{k-1,t}^a$ ($t\in\{1,\cdots,n\}$);
\EndFor
\end{algorithmic}
\end{algorithm}

\subsection{NUISE Algorithm Derivation}\label{sec:AP_n}
%NUISE is the essential part of the robot intrusion detection algorithm~\eqref{nalgo1} where it is recursively called.
%The NISE takes inputs: on-board sensor outputs $\textbf{z}_k$, control input $\textbf{u}_k$, previous state estimate $\hat{\textbf{x}}_{k-1|k-1}$, previous error covariance $P_{k-1}^x$.
%We present the derivation of the NUISE algorithm.
%The NUISE produces state and actuator attack estimates, and their error covariance matices where the error covariance represents how the estimated state is accurate. 
%Two important criteria are applied to the derivation of the NUISE. That is, the estimate is unbiased and its error covariance is minimized. We use a linear estimator gain called the best linear unbiased estimator (BLUE)
Minimum variance unbiased state and unknown input estimation is first introduced in~\cite{kitanidis1987unbiased} with indirect feedthrough only. This result is extended by many research.
A general parameterized gain matrix is derived in~\cite{darouach1997unbiased}, and direct feedthrough unknown input estimation is integrated into the system in~\cite{cheng2009unbiased,hou1998optimal}.
Paper~\cite{SY-MZ-EF:Automatica16} analyze
the stability of the system with direct and indirect feedthrough unknown input.
%by generalizing the solution~\cite{darouach1997unbiased}, integrating directly measurable unknown input~\cite{cheng2009unbiased,hou1998optimal}, and analyzing the stability~\cite{SY-MZ-EF:Automatica16}.
The estimator with indirect feedthrough unknown input has been applied to system fault detection~\cite{chen1996design} without noise and~\cite{de1997nonparametric,liu2011robust} with noise.
The estimator with direct and indirect feedthrough unknown input is applied to attack detection~\cite{yong2016simultaneous} with noise in which the attack location is unknown. However, all the current research is limited to linear dynamic systems.
The proposed NUISE is an extension of the above references to nonlinear systems. It is also an extension of 
the extended Kalman filters~\cite{jazwinski2007stochastic} for state estimation of nonlinear systems by integrating unknown input estimation.
This is the first time to study the state and unknown input estimation on a class of stochastic nonlinear systems.

To find an optimal estimate, we first define what is the meaning of being optimal.
The optimality contains two properties. Firstly, the estimate is unbiased; i.e., its expected value is equal to the targeted value. Secondly, the estimate has the minimum error covariance matrix; i.e., estimation error variance must be minimized given information.
%We call such optimal estimate as the best linear unbiased estimator (BLUE).

%To find an optimal estimate, we first define what is the {\color{red}meaning of optimal}. Note that, to be an accurate estimate, its expected value must be the targeted value (unbiased), and its error variance must be smaller than other estimate (minimum variance). Such intuition leads us to the best linear unbiased estimator (BLUE); i.e., unbiased estimate with minimized error covariance matrix. Linear estimation refers that the estimate is obtained by a weighted average of each sensors. Therefore, {\color{red}to find the optimal estimate at each step,} it is essential to keep track of the estimate as well as its error covariance matrice for the BLUE.

We will derive the NUISE through 4 steps: 1) actuator attack estimation, 2) state prediction, 3) state estimation, 4) sensor attack estimation. In each intermediate step, estimation error and covariance matrix are calculated to find the optimal estimates.

Consider the following system which contains~\eqref{ie019} as a special case:
\begin{align}
\textbf{x}_{k+1} &= f_k^j(\textbf{x}_k,\textbf{u}_k+\textbf{d}_{k}^{a,j})+\zeta_k^j\nonumber\\
\textbf{z}_{1,k}^j&=h_{1,k}^j(\textbf{x}_k)+\textbf{d}_{k}^{s,j}+\xi_{1,k}^j\nonumber\\
\textbf{z}_{2,k}^j&= h_{2,k}^j(\textbf{x}_k)+\xi_{2,k}^j
\label{CD001.1}
\end{align}
where attack $\textbf{d}_{k}^{s,j}$ and $\textbf{d}_{k}^{a,j}$ represent sensor attack and actuator attack.
Testing sensor readings $\textbf{z}_{1,k}^j$ might be modified by attack vector $\textbf{d}_{k}^{s,j}$. Reference sensor readings $\textbf{z}_{2,k}^j$ is assumed to be clean in mode $j$.
We omit mode index $j$ in the NISE derivation because each NUISE is associated with fixed $j$.
Dynamic system~\eqref{CD001.1} can be linearized into
\begin{align}
\textbf{x}_{k+1} & \simeq  A_k \textbf{x}_k+ B_k \textbf{u}_k+G_{k}\textbf{d}_{k}^a +\zeta_k\nonumber\\
\textbf{z}_{1,k} & \simeq  C_{1,k} \textbf{x}_k + \textbf{d}_{k}^s+\xi_{1,k}\nonumber\\
\textbf{z}_{2,k} & \simeq  C_{2,k} \textbf{x}_k + \xi_{2,k}
\label{CD001.7}
\end{align} 
where 
\begin{align*}
&A_k \triangleq \frac{\partial f_k}{\partial \textbf{x}}\big|_{\hat{\textbf{x}}_{k|k},\textbf{u}_{k}+\hat{d}_{k-1}^a},
B_k \triangleq \frac{\partial f_k}{\partial u}\big|_{\hat{\textbf{x}}_{k|k},\textbf{u}_{k}+\hat{d}_{k-1}^a},\nonumber\\
&C_{1,k} \triangleq \frac{\partial h_{1,k}}{\partial \textbf{x}}\big|_{\hat{\textbf{x}}_{k|k-1}},
C_{2,k} \triangleq \frac{\partial h_{2,k}}{\partial \textbf{x}}\big|_{\hat{\textbf{x}}_{k|k-1}}\nonumber\\
&G_{k} \triangleq \frac{\partial f_k}{\partial \textbf{d}^a}\big|_{\hat{\textbf{x}}_{k|k},\textbf{u}_{k}+\hat{d}_{k-1}^a}.
\end{align*}

\textbf{Attack $\textbf{d}_{k-1}^{a}$ estimation}: 
Given unbiased previous state estimate $\hat{\textbf{x}}_{k-1|k-1}$, we can predict the current state using the known kinematic function $f_k(\cdot)$ as follows
\begin{align*}
\hat{\textbf{x}}_{k|k-1}^{*} = f_{k-1}(\hat{\textbf{x}}_{k-1|k-1},\textbf{u}_{k-1}).
\end{align*}
The estimate error is described by
\begin{align*}
\tilde{\textbf{x}}_{k|k-1}^* &= \textbf{x}_k-\hat{\textbf{x}}_{k|k-1}^*= A_{k-1} \tilde{\textbf{x}}_{k-1|k-1} +G_{k-1} \textbf{d}_{k-1}^a+ \nonumber\\&\zeta_{k-1}.
\end{align*}
Noticeably, the estimation is biased, i.e., ${\mathbb E}[\hat{\textbf{x}}_{k|k-1}^*] \neq \textbf{x}_{k|k-1}$ because we do not consider possible unknown attacks yet $G_{k-1} \textbf{d}_{k-1}^a \neq 0$.
To have the unbiased state prediction, it is now needed to find the estimate of actuator attack. The expected output without considering the actuator attack will be $C_{2,k}\hat{\textbf{x}}_{k|k-1}^*$. The informational discrepancy between what we expected and what we actually obtain $\textbf{z}_{2,k}-C_{2,k}\hat{\textbf{x}}_{k|k-1}^*$ shows us the effect of attack $\textbf{d}_{k-1}^a$ and thus this term is used to estimate it.
Actuator attacks are estimated linearly from sensor output bias
\begin{align*}
\hat{\textbf{d}}_{k-1}^a &= M_{2,k}(\textbf{z}_{2,k}-C_{2,k}f_{k-1}(\hat{\textbf{x}}_{k-1|k-1},\textbf{u}_{k-1}))\nonumber\\
&=M_{2,k}(C_{2,k} A_{k-1} \tilde{\textbf{x}}_{k-1|k-1}+C_{2,k} G_{k-1}\textbf{d}_{k-1}^a\nonumber\\&+C_{2,k}\zeta_{k-1}+\xi_{2,k})
\end{align*}
where the estimator gain $M_{2,k}$ represents a weight average of sensor bias based on the trustfulness of each sensor.
The unknown input estimate is unbiased, i.e.,
${\mathbb E}[\hat{\textbf{d}}_{k-1}^a]=\textbf{d}_{k-1}^a$ if ${\mathbb E}[\tilde{\textbf{x}}_{k-1|k-1}]=0$, and $M_{2,k}C_{2,k}G_{k-1}=I$.
In order to achieve optimal estimates, matrix gain $M_k$ should be carefully chosen with minimum variance. To do this, consider the sensor output bias
\begin{align*}
\tilde{\textbf{z}}_{2,k} &= C_{2,k}(G_{k-1} \textbf{d}_{k-1}^a + A_{k-1}\tilde{\textbf{x}}_{k-1|k-1}+\zeta_k)+\xi_k
\end{align*}
where ${\mathbb E}[C_{2,k}A_{k-1}\tilde{\textbf{x}}_{k-1|k-1}+C_{2,k}\zeta_{k-1}+\xi_k]=0$ and its covariance is calculated by
\begin{align*}
\tilde{R}_{2,k}^{*} &\triangleq {\mathbb E}[\tilde{\textbf{z}}_{2,k}\tilde{\textbf{z}}_{2,k}^T]=C_{2,k} \tilde{P}_{k-1}C_{2,k}^T+R_{2,k}
\end{align*}
where $\tilde{P}_k \triangleq A_{k-1} P_{k-1}^xA_{k-1}^T+Q_{k-1}$.
%which is positive definite since $P_{k-1}^x$ and $R_k$ are positive definite. Also, $Q_k$ is positive semi-definite. 
We choose the matrix $M_k$ using the Gauss Markov theorem~\cite{kailath2000linear} as
\begin{align*}
M_{2,k}=(G_{k-1}^TC_{2,k}^T\tilde{R}_{2,k}^{*-1}C_{2,k}G_{k-1})^{-1}G_{k-1}^TC_{2,k}^T\tilde{R}_{2,k}^{*-1}
\end{align*}
%Since the Gauss Markov theorem requires normalized noise covariance, we scale the sensor output bias $\tilde{\textbf{z}}_k$ by $S_k^{-1}$ as follows \begin{align*} S_k^{-1}\tilde{\textbf{z}}_k &= S_k^{-1}C_kG_{k-1} \textbf{d}_{k-1} \nonumber\\ &+ S_k^{-1}(C_kA_{k-1}\tilde{\textbf{x}}_{k-1|k-1}+C_k\zeta_k+\xi_k) \end{align*} {\color{red}where the invertible matrix $S_k$ is found from $\tilde{R}_k=S_kS_k^T$.} Using Gauss Markov theorem~\cite{kailath2000linear}, a BLUE can be found by
which satisfies $M_{2,k}C_{2,k}G_{k-1}=I$. We assume that $G_{k-1}^TC_{2,k}^T\tilde{R}_{2,k}^{*-1}C_{2,k}G_{k-1}$ is invertible. Attack vector estimation error covariance is
$P_{k-1}^{a} \triangleq {\mathbb E}[\tilde{\textbf{d}}_{k-1}^a(\tilde{\textbf{d}}_{k-1}^a)^T]=M_{2,k}\tilde{R}_{2,k}^*M_{2,k}^T$.

\textbf{State prediction}: Estimate $\hat{\textbf{x}}_{k|k-1}^*$ was calculated under a partial knowledge on the attack. Since now we have the estimate $\textbf{d}_{k-1}^a$ of attack, we can update the state estimate
\begin{align*}
\hat{\textbf{x}}_{k|k-1}&=f_{k-1}(\hat{\textbf{x}}_{k-1|k-1},\textbf{u}_{k-1}+\hat{\textbf{d}}_{k-1}^a)
\end{align*}
and it is now unbiased ${\mathbb E}[\hat{\textbf{x}}_{k|k}]=\textbf{x}_k$ since ${\mathbb E}[\hat{\textbf{d}}_{k-1}^a]=\textbf{d}_{k-1}^a$.
For the next step, we find the state prediction error covariance matrix
\begin{align}
P_{k|k-1}^x&=\bar{A}_{k-1}P_{k-1}^x\bar{A}_{k-1}^T+\bar{Q}_{k-1}
\label{P_kk_1}
\end{align}
where $\bar{A}_{k-1}=(I-G_{k-1}M_{2,k}C_{2,k})A_{k-1}$
and $\bar{Q}_{k-1}^j = (I-G_{k-1}M_{2,k}C_{2,k})Q_{k-1}(I-G_{k-1}M_{2,k}C_{2,k})^T+G_{k-1}M_{2,k}R_{2,k}M_{2,k}^TG_{k-1}^T$.

\textbf{State estimation}: 
Predicted state $\hat{\textbf{x}}_{k|k-1}$ is not perfect because of process and measurement errors. To have the estimate more accurate, we correct the state estimate using sensor readings. Here, we again utilize the difference between the newly predicted output $C_{2,k}\hat{\textbf{x}}_{k|k-1}$ and real sensor output $\textbf{z}_{2,k}$ to reflect the effect of unknown noises:
\begin{align*}
\hat{\textbf{x}}_{k|k}&=\hat{\textbf{x}}_{k|k-1}+L_{k}(\textbf{z}_{2,k}-h_{2,k}(\hat{\textbf{x}}_{k|k-1}))
\end{align*}
where the state estimate is unbiased ${\mathbb E}[\hat{\textbf{x}}_{k|k}]=\textbf{x}_k$ and the estimate gain matrix $L_k$ will be chosen such that the new estimate $\hat{\textbf{x}}_{k|k}$ has a smaller error variance.
Error dynamic and covariance are
\begin{align*}
\tilde{\textbf{x}}_{k|k}=\textbf{x}_k-\hat{\textbf{x}}_{k|k}= (I-L_{k}C_{2,k})\tilde{\textbf{x}}_{k|k-1}-L_k\xi_{2,k}
\end{align*}
and
\begin{align*}
P_{k}^x&=(I-L_kC_{2,k})P_{k|k-1}^x(I-L_kC_{2,k})^T + L_k R_{2,k} L_k^T\nonumber\\
&-(I-L_k C_{2,k}) G_{k-1}M_{2,k}R_{2,k}L_k^T\nonumber\\
&-L_kR_{2,k}M_{2,k}^TG_{k-1}^T(I-L_k C_{2,k})^T.
\end{align*}
To achieve the optimal estimates, %the minimum covariance estimate, we use the trace norm of $P_{k}^x$ as an objective function of its 
we solve the variance minimization program: $\min_{L_k} {\rm tr} (P_{k}^x)$.
%\begin{align*} \min_{L_k} {\rm tr} (P_{k}^x)%=\min_{L_k} {\rm tr} (P_{k}^{x*}+L_k\tilde{R}_k^*L_k^T-2L_kJ_k^T). \end{align*} This is a non-constrained convex optimization problem. Therefore, the gradient of the objective function will vanish at the minimizer.
We can take derivative the objective function with respect to the decision variable $L_k$ and set it to zero to find the solution:
%\begin{align*} \frac{\partial {\rm tr} (P_{k}^x)}{\partial L_k}=2\tilde{R}_k^*L_k^T-2J_k^T=0. \end{align*} The solution is
$L_k=(C_{2,k}P_{k|k-1}+R_{2,k}M_{2,k}^TG_{2,k-1}^T)^T\tilde{R}_{2,k}^{-1}$
where $\tilde{R}_{2,k} \triangleq C_{2,k}P_{k|k-1}^xC_{2,k}^T+R_{2,k} + C_{2,k}G_{k-1}M_{2,k}R_{2,k}+R_{2,k}M_{2,k}^TG_{k-1}^TC_{2,k}^T$ must be invertible.

\textbf{Attack $\textbf{d}_{k}^{s,j}$ estimation}:
Given $\hat{\textbf{x}}_{k|k}$, the linear estimation for unknown input $\textbf{d}_{k}^s$ can be
\begin{align}
\hat{\textbf{d}}_{k}^s&= M_{1,k} (\textbf{z}_{1,k}- h_{1,k}(\hat{\textbf{x}}_{k|k}))\nonumber\\
&=M_{1,k}(C_{1,k} \tilde{\textbf{x}}_{k|k}+\textbf{d}_{k}^s +\xi_{1,k})
\label{P205}
\end{align}
where the estimate is unbiased ${\mathbb E}[\hat{\textbf{d}}_{k}^s]=\textbf{d}_{k}^s$ if $M_{1,k}=I$. This also can be found by Gauss Markov theorem. By the theorem, the optimal estimate is 
\begin{align*}
M_{1,k} \triangleq ( \tilde{R}_{1,k}^{-1})^{-1}\tilde{R}_{1,k}^{-1}=I
\end{align*}
where $\tilde{R}_{1,k}=C_{1,k}P_{k}^x C_{1,k}^T + R_{1,k}$. Covariance matrices are found by
\begin{align*}
P_{k}^{s}&=\tilde{R}_{1,k}
%\nonumber\\P_{k}^{xd_1}&=(P_{k}^{d_1x})^T=P_{k}^x C_{1,k}^TM_{1,k}^T.
%\label{P207}
\end{align*}

\textbf{Likelihood of the mode}:
It is natural that the predicted output must be matched with the measured output if the mode $j$ is the true mode.
For $\forall j$, we quantify the discrepancy between the predicted output and the measured output as follows
\begin{align*}
\nu_k^j = \textbf{z}_{2,k}-h_{2,k}^j(\hat{\textbf{x}}_{k|k-1}^j)).
\end{align*}
We approximate the output error $\nu_k^{j}$ as a multivariate Gaussian random variable. Then, the likelihood function is given by
\begin{align*}
{\mathcal N}_k^{j}
&\triangleq {\mathcal P}(y_k|j={\rm true})=
{\mathcal N}(\nu_k^{j};0,\bar{P}_{k|k-1}^{j})\nonumber\\
&=\frac{\exp(-(\nu_k^{j})^T(\bar{P}_{k|k-1}^{j})^{\dagger}\nu_k^{j}/2)}{(2 \pi)^{n^{j}/2} |\bar{P}_{k|k-1}^{j}|_{+}^{1/2}}
\end{align*}
where $\bar{P}_{k|k-1}^j = C_{2,k}^jP_{k|k-1}^{x,j}(C_{2,k}^j)^T + R_{2,k}^j-C_{2,k}^jG_{k-1}^jM_{2,k}^jR_{2,k}^j-R_{2,k}^j(M_{2,k}^j)^T(G_{k-1}^j)^T(C_{2,k}^j)^T$ is the error covariance matrix of $\nu_k^j$ and $n^j = Rank(\bar{P}_{k|k-1}^j)$.
Notations $\dagger$ and $|\cdot|_{+}$ refer pseudoinverse and pseudodeterminant, respectively.
By the Bayes' theorem, the a posteriori probability is
$\mu_k^{j} \triangleq {\mathcal P}(j={\rm true}|y_k,\cdots,y_0) =\frac{{\mathcal P}(y_k|j={\rm true}){\mathcal P}(j={\rm true}|y_{k-1},\cdots,y_0)}{\sum_{i=1}^{\mathcal M}{\mathcal P}(y_k|j={\rm true}){\mathcal P}(j={\rm true}|y_{k-1},\cdots,y_0)}=\frac{{\mathcal N}_k^j\mu_{k-1}^j}{\sum_{i=1}^{\mathcal M}{\mathcal N}_k^j\mu_{k-1}^j}$.
However, such update might allow that some $\mu_k^j$ converge to zero.
To prevent this, we modify the posterior probability update to
\begin{align*}
\bar{\mu}_k^{j} = \frac{\mu_k^{j}}{\sum_{i=1}^{\mathcal M}\mu_k^i},
\end{align*}
where $\mu_k^{j} = \max\{{\mathcal N}_k^j\mu_{k-1}^{j},\epsilon\}$ and
$\epsilon>0$ is a pre-selected small constant preventing the vanishment of the mode probability. 
The last step is to generate the state, attack vector, and mode estimates of the maximum a posteriori mode.

Algorithm \ref{nalgo2_full} shows the complete nonlinear unknown input and state estimation algorithm.
\begin{algorithm} \caption{Nonlinear Unknown Input and State Estimation Algorithm (NUISE)} \label{nalgo2_full}
\begin{algorithmic}[1]
\Require $\textbf{u}_{k-1}$, $\hat{\textbf{x}}_{k-1|k-1}$, $\textbf{z}_{1,k}^j$, $\textbf{z}_{2,k}^j$, $P_{k-1}^x$
\Ensure $\hat{\textbf{x}}_{k|k}^j$, $\hat{\textbf{d}}_{k}^{s,j}$, $\hat{\textbf{d}}_{k-1}^{a,j}$, $P_{k}^{x,j}$, $P_{k}^{s,j}$, $P_{k-1}^{a,j}$ ${\mathcal N}_k^j$
%\State Initialize: $\hat{x}_{0|0}={\mathbb E}[x_0]$; $\hat{d}_{1,0} = H_{0,k}^{-1}(y_0-C_0\hat{x}_{0|0}-h_0(u_0,0))$; $P_0^x={\mathbb E}[P_0]$; $P_0^{d_1}=(H_{1,k}^T(C_{1,k}P_{k|k}^x C_{1,k}^T + R_{1,k})^{-1}H_{1,k})^{-1}$; $P_0^{d_1x}=C_{1,0} P_{0|0}^x$;
\State Initialize;
\State $\tilde{P}_{k-1}^j \leftarrow A_{k-1}^j P_{k-1}^x(A_{k-1}^j)^T+Q_{k-1}^j$;
\State $\tilde{R}_{2,k}^{*,j} \leftarrow C_{2,k}^j \tilde{P}_{k-1}^j(C_{2,k}^j)^T+R_{2,k}^j$;
\State $M_{2,k}^j\leftarrow((G_{k-1}^j)^T(C_{2,k}^j)^T(\tilde{R}_{2,k}^{*,j})^{-1}C_{2,k}^jG_{k-1}^j)^{-1}$ $(G_{k-1}^j)^T(C_{2,k}^j)^T(\tilde{R}_{2,k}^{*,j})^{-1}$;
\State $\hat{\textbf{d}}_{k-1}^{a,j} \leftarrow M_{2,k}^j(\textbf{z}_{2,k}^j-C_{2,k}^jf(\hat{\textbf{x}}_{k-1|k-1},\textbf{u}_{k-1}))$;
\State $P_{k-1}^{a,j}\leftarrow M_{2,k}^j\tilde{R}_{2,k}^{*,j}(M_{2,k}^j)^T$;
\State $\hat{\textbf{x}}_{k|k-1}^j \leftarrow f(\hat{\textbf{x}}_{k-1|k-1},\textbf{u}_{k-1}+\hat{\textbf{d}}_{k-1}^{a,j})$;
\State $\bar{A}_{k-1}^j\leftarrow (I-G_{k-1}^jM_{2,k}^jC_{2,k}^j) A_{k-1}^j$;
\State %$\bar{Q}_{k-1} \triangleq {\mathbb E}[\bar{\zeta}_{k-1}\bar{\zeta}_{k-1}^T]$ with $\bar{\zeta}_{k-1} \triangleq -(I-G_{2,k-1}M_{2,k}C_{2,k})(G_{1,k-1}M_{1,k-1}\xi_{1,k-1}+\zeta_{k-1})-G_{2,k-1}M_{2,k}\xi_{2,k}$;
$\bar{Q}_{k-1}^j \leftarrow (I-G_{k-1}^jM_{2,k}^jC_{2,k}^j)Q_{k-1}^j(I-G_{k-1}^jM_{2,k}^jC_{2,k}^j)^T+G_{k-1}^jM_{2,k}^jR_{2,k}^j(M_{2,k}^j)^T(G_{k-1}^j)^T$;
\State $P_{k|k-1}^{x,j}\leftarrow\bar{A}_{k-1}^jP_{k-1}^{x}(\bar{A}_{k-1}^j)^T+\bar{Q}_{k-1}^j$;
%$+\bar{A}_{k-1}R_{2,k}M_{2,k}^TG_{2,k-1}^T+G_{2,k-1}M_{2,k}R_{2,k}\bar{A}_{k-1}^T$;
\State $\tilde{R}_{2,k}^j \leftarrow C_{2,k}^jP_{k|k-1}^{x,j}(C_{2,k}^j)^T+R_{2,k}^j + C_{2,k}^jG_{k-1}^jM_{2,k}^jR_{2,k}^j+R_{2,k}^j(M_{2,k}^j)^T(G_{k-1}^j)^T(C_{2,k}^j)^T$;
\State $L_k^j\leftarrow(C_{2,k}^jP_{k|k-1}^{x,j}+R_{2,k}^j(M_{2,k}^j)^T(G_{k-1}^j)^T)^T(\tilde{R}_{2,k}^j)^{-1}$;
%\State $L_k=P_{k|k-1}^xC_k^T \tilde{R}_{2,k}^{-1}$ where $\tilde{R}_{2,k} \triangleq C_{2,k}P_{k|k-1}^xC_{2,k}^T+R_{2,k}$;
\State $\hat{\textbf{x}}_{k|k}^j \leftarrow \hat{\textbf{x}}_{k|k-1}^j+L_k^j (\textbf{z}_{2,k}^j-h_{2}^j(\hat{\textbf{x}}_{k|k-1}^j))$;
\State $P_{k}^{x,j}\leftarrow(I-L_k^jC_{2,k}^j)P_{k|k-1}^{x,j}(I-L_k^jC_{2,k}^j)^T + L_k^j R_{2,k}^j (L_k^j)^T-(I-L_k^j C_{2,k}^j) G_{k-1}^jM_{2,k}^jR_{2,k}^j(L_k^j)^T-L_k^jR_{2,k}^j(M_{2,k}^j)^T(G_{k-1}^j)^T(I-L_k^j C_{2,k}^j)^T$;
\State $\hat{\textbf{d}}_{k}^{s,j}\leftarrow \textbf{z}_{1,k}^j-h_{1}^j(\hat{\textbf{x}}_{k|k}^j)$; 
\State $P_{k}^{s,j}\leftarrow C_{1,k}^jP_{k}^{x,j} (C_{1,k}^j)^T + R_{1,k}^j$;
\State $\nu_k^j \leftarrow \textbf{z}_{2,k}^j -h_{2}^j(\hat{\textbf{x}}_{k|k-1}^j)$;
\State $\bar{P}_{k|k-1}^j \leftarrow C_{2,k}^jP_{k|k-1}^{x,j}(C_{2,k}^j)^T + R_{2,k}^j-C_{2,k}^jG_{k-1}^jM_{2,k}^jR_{2,k}^j-R_{2,k}^j(M_{2,k}^j)^T(G_{k-1}^j)^T(C_{2,k}^j)^T$;
\State $n^j\leftarrow rank(\bar{P}_{k|k-1}^j)$;
\State ${\mathcal N}_k^j \leftarrow \frac{1}{(2 \pi)^{n^j/2} |\bar{P}_{k|k-1}^j|_{+}^{1/2}}\exp(-\frac{(\nu_k^j)^T(\bar{P}_{k|k-1}^j)^{\dagger}\nu_k^j}{2})$;\protect\footnotemark
\end{algorithmic}
\end{algorithm}
\footnotetext{Notations $\dagger$ and $|\cdot|_{+}$ refer pseudoinverse and pseudodeterminant, respectively. $n^j$ refers to the rank of $\bar{P}_{k|k-1}^j$.}

\subsection{Kinematic Model and Measurement Models}\label{sec:kin_mea_model}
\textbf{Kinematic model} The kinematic model of Khepera includes three states: $(x,y)$ is the robot location at a 2-D plane and $\theta$ is its heading. The control commands are determined by two variables: $v_L$ and $v_R$ are the speeds of the left and right wheels, respectively. %the velocity and angular velocity can be denoted as $v_k=\frac{v_L+v_R}{2}$ and $w_k=\frac{v_R-v_L}{r/2}$, where $r$ is the distance between the two wheels. For deduction simplicity purpose, we use the velocityhttps://www.sharelatex.com/project/5914d5ef91debba170c6fd46 and angular velocity as the inputs, and convert them back to $v_L$ and $v_R$ afterwards. 
Considered with actuator attack $\textbf{d}_{k-1}^a=[d_{k-1}^{a, L}, d_{k-1}^{a, R}]^T$ on the left and right wheel, the kinematic model can be presented as:
\begin{align}
x_{k} &= x_{k-1} + T\cos \theta_{k-1}(v_L+d_{k-1}^{a, L}+v_R+d_{k-1}^{a, R})/2+\zeta_{k-1}^x \nonumber\\
y_{k} &= y_{k-1} + T\sin \theta_{k-1}(v_L+d_{k-1}^{a, L}+v_R+d_{k-1}^{a, R})/2+\zeta_{k-1}^y \nonumber\\
\theta_{k} &= \theta_{k-1} + T(v_R+d_{k-1}^{a, R}-v_L-d_{k-1}^{a, L})/\frac{D}{2}+\zeta_{k-1}^\theta
\label{ie9}
\end{align}
where $\zeta_{k-1}=[\zeta_{k-1}^x, \zeta_{k-1}^y, \zeta_{k-1}^\theta]^T$ is assumed to be zero mean Gaussian process noises, and $D$ is the distance between the left and right wheel on Khepera. %Vectors $\textbf{x}_k = [x_k, y_k, \theta_k]^T$ and $\textbf{x}_{k+1} = [x_{k+1}, y_{k+1}, \theta_{k+1}]^T$ denote current states (position and orientation) and next states, respectively.% Velocity and angular velocity are denoted as $v_k=\frac{v_L+v_R}{2}$ and $w_k=\frac{v_R-v_L}{r/2}$. We specify $\textbf{u}_k = [v_k, \omega_k]^T$ as inputs, which is directly controllable from the planner.

\begin{figure}[!t]
  \centering
  \includegraphics[width =0.85\linewidth]{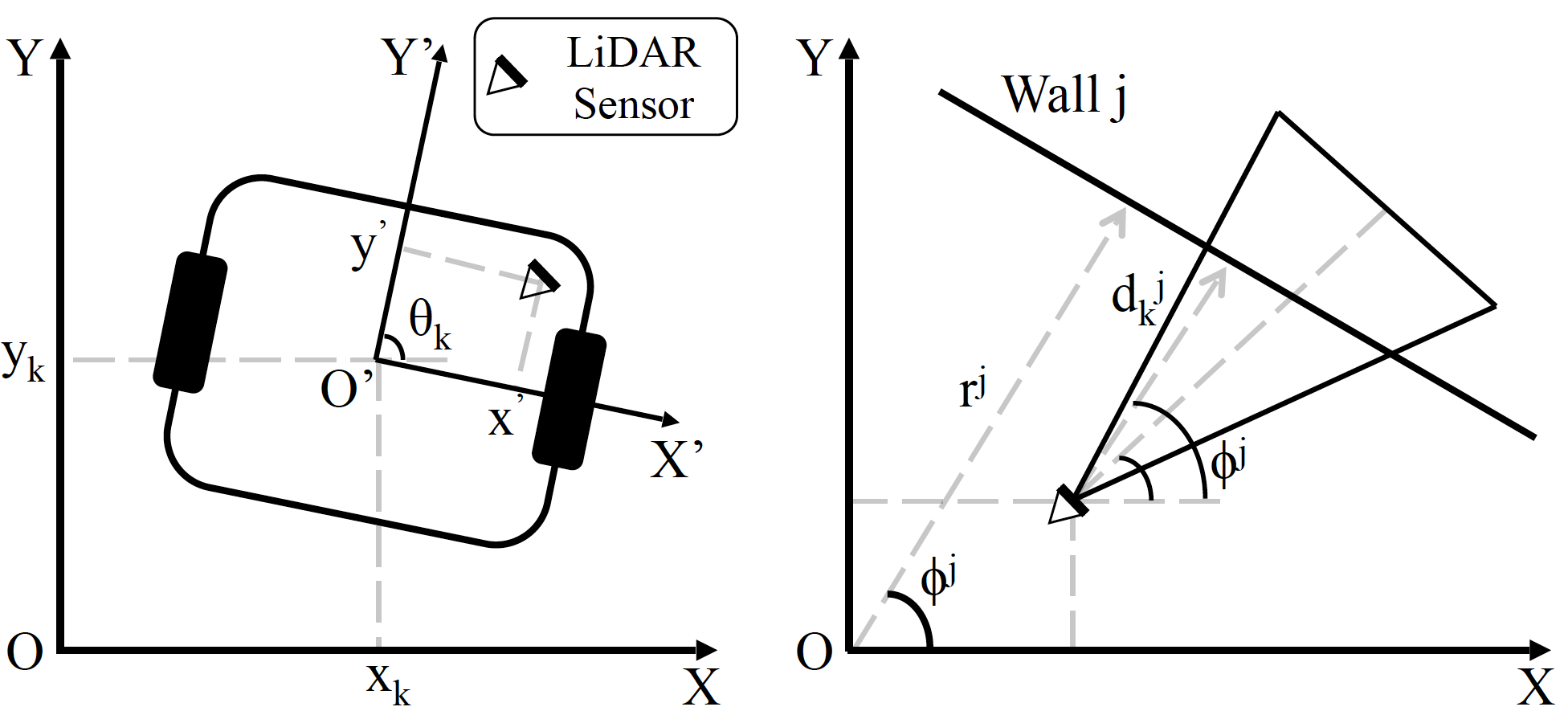}
  \caption{LiDAR sensor measurement model.}\label{khepera_3_math}
\end{figure}

\textbf{Measurement model} The sensor readings include data from three sensors: $\textbf{z}_{k} = [\textbf{z}_{k, I}, \textbf{z}_{k, W}, \textbf{z}_{k, L}]^T$ where $\textbf{z}_{k, I}$ is from IPS, $\textbf{z}_{k, W}$ is from wheel encoder, and $\textbf{z}_{k, L}$ is from LiDAR.

IPS sensor directly measures the states of Khepera, hence, the measurement model can be directly specified by:
\begin{align}
\textbf{z}_{k, I} &= \textbf{x}_{k}+\textbf{d}_{k, I}^s+\xi_{k, I}
\label{eq:measurement_ips}
\end{align}
where $\xi_{k, I}=[\xi_{k, I}^x, \xi_{k, I}^y, \xi_{k, I}^\theta]^T$ refers to measurement noises from IPS and $\textbf{d}_{k, I}^s=[d_{k,I}^{s, x},d_{k,I}^{s, y},d_{k,I}^{s, \theta}]$ refers to the sensor attack vector on IPS. %Suppose $\textbf{x}_{k-1} = [x_{k-1}, y_{k-1}, \theta_{k-1}]^T$ is the previous states of Khepera, the distance measurement from wheel encoder can be referred to as $\textbf{z}_k=[l_L+d_{k, W}^{sL}+\xi_{k, W}^{sL}, l_R+d_{k, W}^{sR}+\xi_{k, W}^{sR}]^T$, where $\xi_{k, W}=[\xi_{k, W}^L, \xi_{k, W}^R]^T$ refers to measurement noises from wheel encoder and $\textbf{d}_{k, W}^s=[d_{k, W}^{sL}, d_{k, W}^{sR}]^T$ refers to the sensor attack vector on the wheel encoder. The measurement model can be represented as:
% \begin{align}
% x_{k} &= x_{k-1} + (z_k(1)+z_k(2))\cos\theta_{k}/2 \nonumber\\
% y_{k} &= y_{k-1} + (z_k(1)+z_k(2))\sin \theta_{k}/2 \nonumber\\
% \theta_{k} &= \theta_{k-1} + (z_k(2)-z_k(1))/r
% \label{eq:measurement_we}
% \end{align}

The raw data measured by the wheel encoder are the distances traveled by each wheel $(l_L, l_R)$. For convenience reason, we convert them into robot states using previous states $\textbf{x}_{k-1}$ before we feed the data to planner:
\begin{align}
x_{k} &= x_{k-1} + (l_L+l_R)\cos\theta_{k}/2 \nonumber\\
y_{k} &= y_{k-1} + (l_L+l_R)\sin \theta_{k}/2 \nonumber\\
\theta_{k} &= \theta_{k-1} + (l_R-l_L)/r \nonumber
\end{align}
Analogously with IPS, the measurement model for the wheel encoder is specified as:
\begin{align}
\textbf{z}_{k, W} &= \textbf{x}_{k}+\textbf{d}_{k, W}^s+\xi_{k, W}
\label{eq:measurement_ips}
\end{align}
where $\xi_{k, W}=[\xi_{k, W}^x, \xi_{k, W}^y, \xi_{k, W}^\theta]^T$ refers to measurement noises from the wheel encoder and $\textbf{d}_{k, W}^s=[d_{k,W}^{s, x},d_{k,W}^{s, y},d_{k,W}^{s, \theta}]^T$ refers to the sensor attack vector on the wheel encoder.

The LiDAR sensor is placed on top of the robot with a shift distance of $[x',y']^T$ from the origin $O'$ as shown in the left of Figure~\ref{khepera_3_math}. Raw sensor readings returned from LiDAR are the distances between LiDAR and surrounding walls and obstacles (see Figure~\ref{fig:env}). Given the LiDAR readings, we process the raw data into the perpendicular distance $l^{j}_k$ from each boundary wall $j \in \{1, 2, 3, 4\}$ and the orientation $\theta_{k, L}$of Khepera. Specifically, we recognize the straight line segments using raw distances from all direction, and calculate the distances to each wall as follows: 
\begin{align}
  l^{j}_k %&= r^j - x_k \cos \phi^j - y_k \sin \phi^j+ \xi_k \nonumber\\
  & =  r^j - (x_k+ x'\sin \theta_k + y' \cos \theta_k) \cos \phi^j\nonumber\\
  & \ \ - (y_k - x'\cos \theta_k + y' \sin \theta_k) \sin \phi^j +d_{k,L}^{s,j} +\xi_{k, L}^j
  \label{ee001}
\end{align}
where $\xi_{k, L}=[\xi_{k, I}^j]^T,j \in \{1, 2, 3, 4\}$ refers to measurement noises from LiDAR. Distance $r^j$ and angle $\phi^j$ for each wall is known in advance as the map information. Using $\phi^j$ of each wall and the 240 degrees of range, we can also infer the angle of the robot. We use the distances to each wall and the angle as the readings from LiDAR: $\textbf{z}_{k, L} =[l_{k,}^{j}, \theta_k]^T,j \in \{1, 2, 3, 4\} $. In outdoor environments, LiDAR measurement model can be obtained using more complicated simultaneous localization and mapping (SLAM) algorithms~\cite{durrant2006simultaneous}. For demonstration purpose, we apply a simple transformation in the indoor environment~\cite{jetto1999development}.

\subsection{Separating Actuator Attack Vector}\label{sec:separation}
%correlated covariance problem
Without loss of generosity, we consider a robot with two actuators such as Khepera. During actuator attack vector estimation, we obtain $\hat{\textbf{d}}_k^a =[\hat{d}_k^L, \hat{d}_k^R]^T$, with error covariance $P_k^a$. In Algorithm~\ref{nalgo1}, we test 
\begin{align}
(\hat{\textbf{d}}_k^a)^T (P_k^a)^{-1} \hat{\textbf{d}}_k^a \geq \chi_{p=2}(\alpha)
\label{ttp0}
\end{align}
to determine the existence of actuator attacks. Threshold $\chi_{p=2}(\alpha)$ is a Chi-square test value with degree of freedom $p=2$ and confidence level $\alpha$.

In order to confirm actuator attack on each actuator, we need to separately conduct Chi-square test $\hat{d}_k^L$, and $\hat{d}_k^R$, with corresponding marginal variances $P_k^a(1,1)$, and $P_k^a(2,2)$:
\begin{align}
(\hat{d}_k^L)^T (P_k^a(1,1))^{-1} \hat{d}_k^L \geq \chi_{p=1}^2(\alpha)\nonumber\\
(\hat{d}_k^R)^T (P_k^a(2,2))^{-1} \hat{d}_k^R \geq \chi_{p=1}^2(\alpha).
\label{ttp1}
\end{align}
However, a positive testing result in~\eqref{ttp0} does not guarantee a positive testing result in~\eqref{ttp1} because off-diagonal terms of matrix $P_k^a$ are neglected in~\eqref{ttp1}. The explanation is shown as follows:
\begin{align}
&(\hat{\textbf{d}}_k^a)^T (P_k^a)^{-1} \hat{\textbf{d}}_k^a = (\hat{d}_k^L)^T (P_k^a)^{-1}(1,1) \hat{d}_k^L\nonumber\\
&\quad \quad \quad +(\hat{d}_k^L)^T (P_k^a)^{-1}(1,2) \hat{d}_k^R
+(\hat{d}_k^R)^T (P_k^a)^{-1}(2,1) \hat{d}_k^L\nonumber\\
&\quad \quad \quad +(\hat{d}_k^R)^T (P_k^a)^{-1}(2,2) \hat{d}_k^R\nonumber\\
&(\hat{d}_k^L)^T (P_k^a(1,1))^{-1} \hat{d}_k^L = (\hat{d}_k^L)^T (P_k^a(1,1))^{-1} \hat{d}_k^L\nonumber\\
&(\hat{d}_k^R)^T (P_k^a(2,2))^{-1} \hat{d}_k^R = (\hat{d}_k^R)^T (P_k^a(1,1))^{-1} \hat{d}_k^R
\end{align}
%We cannot compare~\eqref{ttp0} and~\eqref{ttp1} directly since $(P_k^a(1,1))^{-1} \neq (P_k^a)^{-1}(1,1)$.
Note that $(\hat{\textbf{d}}_k^a)^T (P_k^a)^{-1} \hat{\textbf{d}}_k^a=(\hat{d}_k^L)^T (P_k^a(1,1))^{-1} \hat{d}_k^L+(\hat{d}_k^R)^T (P_k^a(2,2))^{-1} \hat{d}_k^R$ if $P_k^a$ is a diagonal matrix.

% nonlinearity of Chi-square threshold
%Even though $P_k^a$ is a diagonal matrix so that we could separate them, there is another problem that Chi-square test threshold is nonlinear; i.e., 
Another problem for the separation is that Chi-square test threshold is nonlinear. For instance, 
$\chi_{p=1}^2(0.01)=6.635$ and $\chi_{p=2}^2(0.01)=9.210$.
Suppose $P_k^a$ is a diagonal matrix and the test scores after separation are $(\hat{d}_k^L)^T (P_k^a(1,1))^{-1} \hat{d}_k^L=5$ and $(\hat{d}_k^R)^T (P_k^a(2,2))^{-1} \hat{d}_k^R=5$. The actuator attack would be detected by~\eqref{ttp0} but not by~\eqref{ttp1}.

Hence, we conduct the Chi-square test for the aggregated actuator attacks.

\subsection{Building RIDS on UAV}\label{sec:app_uav}
To further demonstrate the general applicability of RIDS, we will elaborate how to build an RIDS on UAV. Consider an UAV which is mounted with an inertial navigation system (IMU) and a GPS.
The state of the UAV can be specified as $\textbf{x}_k \triangleq [x_k,y_k,z_k,v_k^x,v_k^y,$ $v_k^z,\phi_k,\theta_k,\psi_k]^T$, which denotes displacements, velocities, and angles on X, Y and Z axises. Inputs $\textbf{u}_k \triangleq [p,q,r,a^x,a^y,a^z]^T$ are rotation rates and accelerations on three axises. The kinematic model of the UAV is given in \eqref{ie019} where the kinematic function $f(\cdot)$ is as follows: 
\begin{align*}  &f(\textbf{x}_k,\textbf{u}_k) =\nonumber\\& \left[ \begin{array}{c} \scriptstyle \left[ \begin{array}{c}x_k\\y_k\\z_k\\\end{array}\right]+T R_k^T \left[ \begin{array}{c}v_k^x\\v_k^y\\v_k^z\\ \end{array}\right]\\ \scriptstyle \left[ \begin{array}{c}v_k^x\\v_k^y\\v_k^z\\\end{array}\right]+T \left[ \begin{array}{c} a_k^x + v_k^y r - v_k^z q + g s(\theta_k)\\ a_k^y - v_k^x r + v_k^z p - g c(\theta_k)s(\phi_k)\\ a_k^z + v_k^x q - v_k^y p - g c(\theta_k)c(\phi_k)\\ \end{array} \right]\\ \scriptstyle \left[ \begin{array}{c}\phi_k\\\theta_k\\\psi_k\\\end{array}\right]+T \left[ \begin{array}{ccc} 1&s(\phi_k)tan(\theta_k)&c(\phi_k)t(\theta_k)\\ 0&c(\phi_k)&-s(\phi_k)\\ 0&s(\phi_k)sec(\theta_k)&c(\phi_k)sec(\theta_k)\\ \end{array} \right] \left[ \begin{array}{c}p_k\\ q_k\\ r_k\\ \end{array}\right] \end{array} \right] \end{align*} where $R_k$ is \begin{equation} \scriptstyle \begin{bmatrix} 1 & 0 & 0\\ 0 & c(\phi_k) & s(\phi_k)\\ 0 & -s(\phi_k)&c(\phi_k)\\ \end{bmatrix} \begin{bmatrix} c(\theta_k)& 0 & -s(\theta_k)\\ 0 & 1 & 0\\ s(\theta_k) & 0 & c(\theta_k)\\ \end{bmatrix} \begin{bmatrix} c(\psi_k) & s(\psi_k) & 0\\ -s(\psi_k) & c(\psi_k) & 0\\ 0 & 0 & 1\\ \end{bmatrix} \notag \end{equation} and $c(\cdot)$, and $s(\cdot)$ refer to $\cos(\cdot)$ and $\sin(\cdot)$ respectively. 
GPS only measures the location of UAV, hence the measurement function is $h(\textbf{x}_k)=[x_k, y_k, z_k]^T$. An IMU generates full states, hence the measurement function can be described as $h(\textbf{x}_k) = \textbf{x}_k$. With the above model, we can apply RIDS to detect the attacks on UAV.

\subsection{Detection Results under Attack Scenarios}\label{sec:more_figures}
More detection results from RIDS under attack scenarios in Table~\ref{tab:attack_setup} are shown in Figure~\ref{fig:appendix_results}. Figure~\ref{fig:est_prob_n6} shows the detection output when there is neither actuator attack nor sensor attack. Estimation results in plot 1-4 show nearly zero attack vector estimates. The Chi-square test statistics shown in plot 5 and 7 indicate both actuator and sensor attack remain under the threshold, except some occasional spikes. After the sliding window filtering, plot 6 and 8 indicates an attack silence.

\end{document}